\documentclass[12pt,draftcls,onecolumn]{IEEEtran}

\usepackage[english]{babel}

\usepackage[cmex10]{amsmath}
\usepackage{epsfig}
\usepackage[cp1250]{inputenc}
\usepackage{amssymb}

\makeatletter
\let\@ORGmakecaption\@makecaption
\long\def\@makecaption#1#2{\@ORGmakecaption{#1}{#2}\vskip\belowcaptionskip\relax}
\makeatother
\usepackage[numbers,square]{natbib}

\newcounter{MYtempeqncnt}

\newcommand{ \pppM }{ \Theta_\text{MBS} }
\newcommand{ \pppF }{ \Theta_\text{FBS} }
\newcommand{ \pppP }{ \Theta_\text{PBS} }
\newcommand{ \pppMUE }{ \Theta_\text{MUE} }
\newcommand{ \intM }{ \lambda_\text{M} }
\newcommand{ \intF }{ \lambda_\text{F} }
\newcommand{ \intP }{ \lambda_\text{P} }
\newcommand{ \intMUE }{ \lambda_\text{MUE} }
\newcommand{ \pM }{ P_\text{M} }
\newcommand{ \pF }{ P_\text{F} }
\newcommand{ \pP }{ P_\text{P} }
\newcommand{ \lM }{ \phi_\text{M} }
\newcommand{ \lF }{ \phi_\text{F} }
\newcommand{ \lP }{ \phi_\text{P} }

\newcommand{ \pl }{ H }
\newcommand{ \ple }{ \alpha }
\newcommand{ \pleM }{ \alpha_{\text{M}} }
\newcommand{ \pleF }{ \alpha_{\text{F}} }
\newcommand{ \pleP }{ \alpha_{\text{P}} }

\newcommand{ \sir }{ \gamma }
\newcommand{ \pr }{ \mathbb{P} }
\newcommand{ \kDI }{ k }
\newcommand{ \ev }{ \mathbb{E} }
\newcommand{ \aRes }{ \rho_{\text{A}} }
\newcommand{ \nAbsf }{ N_{\text{A}} }
\newcommand{ \nSf }{ N_{\text{S}} }
\newcommand{ \nRb }{ N_{\text{R}} }
\newcommand{ \rM }{ r_{\text{M}} }
\newcommand{ \rF }{ r_{\text{F}} }
\newcommand{ \rP }{ r_{\text{P}} }
\newcommand{ \cae }{ \kappa  }
\newcommand{ \mf }{ \varepsilon }
\newcommand{ \nUE }{ L }

\begin{document}

\title{On Number of Almost Blank Subframes in Heterogeneous Cellular Networks}

\author{
Michal~\v{C}ierny,~\IEEEmembership{Student Member,~IEEE,}
Haining~Wang,
Risto~Wichman,
Zhi~Ding,~\IEEEmembership{Fellow,~IEEE,}
Carl~Wijting,~\IEEEmembership{Senior Member,~IEEE}%
\thanks{Manuscript created November 9, 2012; revised April 7 and July 17, 2013; accepted July 29, 2013. This work was supported by Academy of Finland, National Science Foundation grants 1147930 and 1321143, TEKES, Graduate School in Electronics, Telecommunications and Automation (GETA) and HPY Foundation.}%
\thanks{M. \v{C}ierny and R. Wichman are with Department of Signal Processing and Acoustics, Aalto University School of Electrical Engineering, 00076 Aalto, Finland (e-mail: \{michal.cierny, risto.wichman\}@aalto.fi).}
\thanks{H. Wang and Z. Ding are with Department of Electrical and Computer Engineering, University of California, Davis, California 95616 (e-mail: \{hnwang, zding\}@ucdavis.edu).}
\thanks{C. Wijting is with Nokia Research Center in Otaniemi, Otaniementie 19, 02150 Espoo, Finland (e-mail: carl.wijting@nokia.com).}
}

\maketitle

\begin{abstract}
In heterogeneous cellular scenarios with macrocells, femtocells or picocells users may suffer from significant co-channel cross-tier interference. To manage this interference 3GPP introduced almost blank subframe (ABSF), a subframe in which the interferer tier is not allowed to transmit data. Vulnerable users thus get a chance to be scheduled in ABSFs with reduced cross-tier interference. We analyze downlink scenarios using stochastic geometry and formulate a condition for the required number of ABSFs based on base station placement statistics and user throughput requirement. The result is a semi-analytical formula that serves as a good initial estimate and offers an easy way to analyze impact of network parameters. We show that while in macro/femto scenario the residue ABSF interference can be well managed, in macro/pico scenario it affects the number of required ABSFs strongly. The effect of ABSFs is subsequently demonstrated via user throughput simulations. Especially in the macro/pico scenario, we find that using ABSFs is advantageous for the system since victim users no longer suffer from poor performance for the price of relatively small drop in higher throughput percentiles.
\end{abstract}


\section{Introduction}

\IEEEPARstart{A}{lmost} blank subframes (ABSFs) are part of Enhanced Inter-Cell Interference Coordination (eICIC) framework \cite{TS36300} that the 3GPP members have proposed \cite{contr_absf} as means to combat excessive co-channel cross-tier interference in heterogeneous network (HetNet) scenarios. HetNet scenarios are generally cellular network scenarios that cover different types of low-power nodes, such as base stations (BSs), relays or remote radio heads, as underlay to the traditional macrocell tier. HetNet scenarios that are specifically targeted to benefit from ABSFs are combinations of macrocells with closed access femtocells (macro/femto) and macrocells with open access picocells (macro/pico) \cite{PeGuRoKoQuZh2011}.

Femto base station (FBS), also called Home eNodeB (HeNB), is a low-power BS that is deployable by the end user and connects to the core network of a cellular operator by means of wired broadband connection. In the eyes of the operator this is a win-win situation as the users (femto user equipments, FUEs) benefit from higher connection throughput while the use of commonly available wired broadband (such as e.g. digital subscriber line) decreases costs of expanding network infrastructure. In a closed access femtocell only selected users have access to the FBS services, creating thus a closed subscriber group (CSG). A drawback of this is that a non-member macro UE (MUE) that is located close to a closed access FBS can suffer from excessive interference. Proposed (non-ABSF) solutions for downlink interference management in OFDMA femtocells include for example FBS power control \cite{LiQiKa2009}, frequency partitioning \cite{PeVaRoZa2009}, \cite{BhSaAuHa2010}, precoding \cite{ElAhDiLi2012}, cognitive radio approach \cite{LiTsChSu2010} and augmentation of scheduling algorithms \cite{ElAhDiLiHaWi2012}.

Pico base station (PBS) is practically a normal base station with lower transmit power and therefore smaller coverage region. The point of PBS deployment lies not in covering areas where macro tier signal is too low, but in augmenting the macro tier in areas where the concentration of MUEs is too high to be efficiently served by a macro base station (MBS). For such augmentation to be successful it has been shown (see \cite{contr_cre} for the first suggestion) that even UEs that have somehow stronger signal from the closest MBS should be allowed to associate to a PBS, thus leading to a so-called cell range expansion (CRE) concept. Hence, more UEs will associate with PBSs, leading to more efficient frequency reuse and desirable traffic offloading from the macro tier. However, as the CRE description already suggests, some pico UEs (PUEs) see strong interference from the macro tier. Proposed (non-ABSF) solutions for downlink interference management in the macro/pico scenario include interference cancellation \cite{WaDu2011}, frequency partitioning \cite{HuLi2011} and MBS power control \cite{LoCh2011}.

The ABSF concept is based on blanking some subframes of the interferer tier and scheduling the especially vulnerable UEs in these subframes. The vulnerable users thus get part of radio resources where cross-tier interference is lower. The ABSFs are called almost blank because not all resource elements are allowed to be blanked - the cell-specific reference symbols (CRS) that are used for radio resource management (RRM) measurements and channel estimation have to remain present. The strong interference in CRS resource elements is a separate issue and was suggested to be tackled by interference cancellation (see \cite{HuXu2012} or \cite{DaMoChJiYaZo2012} also for other control channel challenges), but such considerations are out of scope of this work. Alternatively to ABSF, the BS can configure an empty MBSFN (Multicast-Broadcast Single Frequency Network) subframe, but its use is more constrained, therefore, our focus will be on the ABSF. Compared to other mentioned interference management solutions, the ABSF concept is simple enough to be incorporated into often technically entangled 3GPP specifications and at the same time it has found rather wide acceptance among the standardization partners.

The interferer tier in the macro/femto scenario is the femto tier, while in the macro/pico scenario it is the macro tier. In case there are both FBSs and PBSs within MBS coverage, there might be need for ABSFs in femto tier as well as in macro tier. As the (significant) cross-tier interference can come from multiple BSs, the amount and position of ABSFs has to be coordinated within the network. Indeed, the organization of ABSFs is planned as a part of self organizing network (SON) concept \cite{HaSaSa2012}. In our work we will propose how the number of ABSFs for downlink interference management can be set globally based on BS placement statistics. Such relationship can then serve as an initial estimate or as a backup solution when the distributed coordination does not serve its purpose. We will derive the necessary number of ABSFs for macro/femto and macro/pico scenarios separately and, if needed, the results can be easily combined.

To our best knowledge the question of deriving the number of necessary ABSFs has not been addressed before this work. Besides the mentioned concept research \cite{PeGuRoKoQuZh2011}, \cite{DaMoChJiYaZo2012} the work is quite sparse. For the macro/femto scenario, some simulation results of using ABSFs have been published in \cite{WaPe2011} and \cite{GhMaRaMoCuViThAnXiJoDhNo2012}. In \cite{KaEl2012} the authors introduce a coordination framework for ABSFs, including channel quality indication (CQI) processing, and suggest control messages that are needed for such operation. Simulated performance of the macro/pico scenario has been shown in \cite{OkNaYaSaKu2011}, while \cite{Gu2011} presents also analytical insight into the topic. During the second round of the review process of our work, a solid article on the topic has become available \cite{SiAn2013}. The authors use similar model as we do and derive a rate coverage of the system. They do not however consider residue interference in the ABSF.

We address the problem by setting the number of ABSFs globally using tools from stochastic geometry \cite{StKeMe1996}, \cite{BaBlMu2009}. Base stations and users are modeled as 2D stochastic processes and spatial relation between a user and its closest interferer is leveraged to define victim users, i.e., users that require interference management. Parametrization of the stochastic models and properties of the victim users are then used to formulate the number of necessary ABSFs. We thus give a semi-closed form connection between the stochastic intensity and other parameters and the minimum number of subframes that can be quickly used to determine the fraction of radio resources needed for interference management. Subsequently, we analyze dependence on individual model parameters, the most important result from which is that in macro/pico scenario the residue ABSF interference has a strong effect on the required number of ABSFs. Finally, we demonstrate the effect of derived number of ABSFs on user throughput via simulations. The results show moderate performance gain for victim users in macro/femto scenario, but in macro/pico scenario the improvement is substantial.

The remainder of the paper is structured as follows: Section \ref{sec:sysMod} introduces the stochastic geometry-based system model and defines victim UEs. In Section \ref{sec_succProb} we derive the success probability of victim UEs, i.e., probability that signal-to-interference ratio (SIR) is higher than a predefined threshold. In Section \ref{sec:numAbsfs} we use the success probability to set condition for the necessary number of ABSFs. In Section \ref{sec_perf} we evaluate the effect of ABSFs by means of Monte Carlo simulations. Section \ref{sec:outro} concludes the paper. Most of the sections are divided into two parts, one for macro/femto scenario and one for macro/pico scenario.

\section{System model}
\label{sec:sysMod}

In this work we model BS and UE placements as homogeneous Poisson point processes (PPP). In \cite{AnBaGa2011} it has been shown that the random BS placement produces a good lower bound for SINR distribution, a regular grid BS deployment gives an upper bound, and the actual truth lies somewhere in between. In \cite{TaDhNoAn2012} the authors have taken two samples of real world BS placements and shown that PPP is not actually a good model for them, because it lacks interaction between points. Nevertheless, the use of PPP model is prevalent (see e.g. \cite{CheQuKo2012} for heterogeneous networks) as it offers a rare analytical insight into a larger scale network.

Radio channel conditions are modeled by a combination of distance dependent path loss $\pl(r) \! = \! r^{-\ple}$, where $\ple$ is path loss exponent, and fast Rayleigh fading with exponential power distribution $h \!\sim\! \text{exp}(1)$. For the sake of tractability we consider two general simplifications. Firstly, we do not model shadow fading. Although it is possible to include shadow fading in initial model equations, a considerable degree of tractability is lost (see \cite{AnBaGa2011} and \cite{Mu2012_2}). Secondly, the residue ABSF interference is considered white. In reality, the victim receiver would see full interference on resource elements where the interferer transmits CRS, and some leakage on other resource elements. However, considering that the CRS position varies between cells and all data is subject to scrambling, the white interference model is not extremely far-fetched.

We note here that although 3GPP specifications and state of the art research work offer quite a few techniques that could supplement the use of ABSF for cross-tier interference management, the scope of this study and structure of the system model does not allow us to consider them. We assume a single antenna transceiver at both BS and UE, hence multi-antenna techniques are not considered. Coordinated multi-point transmission and advanced receiver processing are also out of our scope.

\subsection{Macro/femto scenario}

In the first scenario we have an overlay macrocell PPP $\pppM$ of intensity $\intM$ and an underlay femtocell PPP $\pppF$ of intensity $\intF$. Macro UEs (MUEs) form another PPP $\pppMUE$ of intensity $\intMUE$. Because we are interested in protecting MUEs from CSG femtocells, we do not need to model femto UEs. Processes $\pppM$, $\pppF$ and $\pppMUE$ are all separate and independent among each other. Indoor/outdoor positions and walls are not considered in the model. MUE is associated to a macro BS (MBS) with the best long-term channel conditions, i.e., the geographically closest one, as in \cite{AnBaGa2011}. Transmission powers of MBS and FBS are denoted by $\pM$ and $\pF$, respectively. Path loss exponent on MBS-UE links is $\pleM$, on FBS-UE links it is $\pleF$. Data traffic in macro and femto layer is modeled by base station load values $\lM$ and $\lF$, respectively. A comprehensive analysis of load impact on PPP-based model of cellular network has been done in \cite{DhGaAn2012}. Although our work uses a different UE-BS association model, the traffic model philosophy remains the same.

Let us now have an MUE that is associated to MBS of distance $\rM$. We say that an FBS is a dominant interferer (DI) to given MUE if the MUE-FBS distance $\rF$ is smaller than $\kDI \rM$, where $\kDI$ is a DI-defining coefficient. We then define a victim MUE as an MUE that has one or more DIs. MUE can measure long term channel conditions (and thus estimate distance) of FBSs by performing a reference-signal-received-power (RSRP) measurement. Presence of DIs (or victim status) is then reported to MBS, which uses the knowledge to decide which MUEs will be scheduled in ABSF and which in normal subframe (NSF).

Value of the DI-defining coefficient $\kDI$ is important here. Its basic purpose is to take into account difference between MBS and FBS transmission powers and path loss exponents and consider the maximum interferer power that the MUE receiver can withstand. If, for example, $\pleM \!=\! \pleF \!=\!\ple$ and the minimum required SIR at MUE is $0\text{dB}$, $\kDI \!=\! \left( \pF / \pM \right)^{1 / \ple}$. With $\pleM \!\neq\! \pleF$ the relation between received powers can no longer be transformed to linear relation between $\rF$ and $\rM$ and our condition becomes only approximate.

The definition of victim MUE gives $\kDI$ also another dimension. It might happen that interference from multiple FBSs that are not marked as DIs is unbearable and the given MUE should be marked as victim. A safety margin in $\kDI$ value is an option how to counterweight this issue.

Our definition of DIs via $\kDI$ resembles \cite{GuJeWaIn2008} where a contour of equal power is used to decide how to divide available spectrum. The authors however do not pursue its effect on SIR, nor do they analyze it from statistical point of view.

\subsection{Macro/pico scenario}

In the second scenario there is again an overlay macrocell PPP $\pppM$ of intensity $\intM$ and an underlay picocell PPP $\pppP$ of intensity $\intP$. Macro UEs (MUEs) form again a PPP $\pppMUE$ of intensity $\intMUE$. Processes $\pppM$, $\pppP$ and $\pppMUE$ are separate and independent among each other. As we now have an open access policy, part of the MUEs will actually be associated to PBSs and thus called PUEs. Drawing inspiration from \cite{CheQuKo2012}, an MUE is associated to a PBS if the distance to the closest PBS $\rP$ is smaller than $k_1 \rM$, where $\rM$ is the distance to the closest MBS. In case there is no PBS within $k_1 \rM$ distance, the MUE is associated to the closest MBS. The $k_1$ coefficient takes into account differences in MBS and PBS transmission powers $\pM$ and $\pP$, respectively, and the association bias $\cae$. Unlike \cite{SiDhAn2013} that tries to find a framework for $\cae$ optimization, we keep the value constant and analyze its effect on the required number of ABSFs. Path loss exponent on MBS-UE links is $\pleM$, on PBS-UE links it is $\pleP$. The MBS and PBS load values are denoted as $\lM$ and $\lP$, respectively.

The dominant interferers to a PUE are MBSs that fulfill $\rP \!>\! k_2 \rM$, where $\rP$ is distance between PUE and the associated PBS, $k_2$ is the DI-defining coefficient and $\rM$ is distance to given MBS. With positive association bias we have $k_1 > k_2$. Victim PUEs are PUEs that have one or more DIs. They are thus identified by a pair of inequalities $k_2 \rM \!<\! \rP \!<\! k_1 \rM$. Presence of DIs (or victim status) are reported from PUE to PBS, which can then use the knowledge to request ABSFs from the macro tier. The $k_2$ coefficient has the same meaning as $k$ in macro/femto scenario and can as well be used to address the issue of multiple DIs.

\section{Success probability}
\label{sec_succProb}

In this section we derive success probabilities of victim MUEs and PUEs in our scenarios, which we then use in Section \ref{sec:numAbsfs} to set the necessary number of ABSFs. We use the name success probability as in \cite{CheQuKo2012}, i.e., a probability that UE has a signal-to-interference-ratio (SIR) higher than an outage threshold. SIR is used in this work to approximate signal-to-interference-plus-noise-ratio (SINR), as we are modeling interference limited networks. The success probability represents a CCDF of SIR. While for arbitrary located UEs equivalent result have been presented e.g. in \cite{DhGaBaAn2012}, \cite{JoSaXiAn2011} and \cite{Mu2012_1}, conditioning on presence of dominant interferers has to our best knowledge not been done before.

\subsection{Macro/femto scenario}

Signal-to-interference-ratio at a victim MUE on single resource block is given by formula
\begin{equation}
  \sir = \frac{\pM h \rM^{-\pleM}}{I_\text{M}+\aRes \left( I_\text{DI} + I_\text{F} \right)},
\end{equation}
where $h$ denotes fast fading power gain, $I_\text{M}$ denotes sum interference power from the macro tier (all MBSs except associated one), $I_\text{DI}$ denotes sum interference power from DIs, $I_\text{F}$ denotes sum interference power from non-DI FBSs and $\aRes$ denotes residue ABSF interference. Results in this section are derived for ABSF, for NSF one would simply omit $\aRes$. Success probability is defined as
\begin{equation}
  \pr \left\{ \sir > \sir_0 \right\},
\end{equation}
where $\sir_0$ represents the outage threshold.

To increase clarity and give insight on how the work has progressed we first derive success probability in case of full load and single DI present. After that, we generalize it for arbitrary load values and one or more DIs present.

\subsubsection{A single dominant interferer}

Probability that the number of FBS DIs $N_\text{DI}^{(\text{F})}$ around a randomly chosen MUE equals one is given from the definition of PPP as
\begin{equation}
  \pr \left\{ \left. N_\text{DI}^{(\text{F})} = 1 \right| \rM \right\} = \pi \kDI^2 \intF \rM^2 \exp \left( -\pi \kDI^2 \intF \rM^2 \right).
\end{equation}
By averaging over $\rM$ we get
\begin{align}
  \pr \left\{ N_\text{DI}^{(\text{F})} = 1 \right\} & = \int_{\rM} \pr \left\{ \left. N_\text{DI}^{(\text{F})} = 1 \right| r
    \right\} f_{\rM}(r) dr \\
  & = \int_0^\infty \pr \left\{ \left. N_\text{DI}^{(\text{F})} \right| r \right\} 2 \pi \intM r \nonumber \\
  & \quad \times \exp \left( -\pi \intM r^2 \right) dr \\
  & = \frac{\kDI^2 \intF \intM}{\left( \kDI^2 \intF + \intM \right)^2}.
\end{align}
CDF of $\rM$ conditioned on 1 FBS within $\kDI \rM$ is then
\begin{align}
  F_{\rM | \kDI}^{(1)}(R) & = \pr \left\{ \rM \leq R | N_\text{DI}^{(\text{F})} = 1 \right\} \\
  & = \frac{ \pr \left\{ \rM \leq R , N_\text{DI}^{(\text{F})} = 1 \right\} }
    { \pr \left\{ N_\text{DI}^{(\text{F})} = 1 \right\} } \\
  & = \frac{ \int_0^R \pr \left\{ \left. N_\text{DI}^{(\text{F})} = 1 \right| \rM = r \right\} f_{\rM}(r) dr }
    { \frac{\kDI^2 \intF \intM}{\left( \kDI^2 \intF + \intM \right)^2} } \\
  & = \frac{\left( \kDI^2 \intF + \intM \right)^2}{\kDI^2 \intF \intM} \int_0^R \pr \left\{ \left. N_\text{DI}^{(\text{F})} \!=\! 1 \right| 
    \rM \!=\! r \right\} \nonumber \\
  & \quad \times 2 \pi \intM r \exp \left( \!-\pi \intM r^2 \right) dr \\
  & = 1- \left( 1 + \pi \left( \kDI^2 \intF + \intM \right) R^2 \right) \nonumber \\
  & \quad \times \exp \left( -\pi \left( \kDI^2 \intF + \intM \right) R^2 \right).
\end{align}
By differentiation we get a PDF
\begin{equation}
  \label{eq_rmCond}
  f_{\rM | \kDI}^{(1)}(r) = 2 \pi^2 \left( \kDI^2 \intF + \intM \right)^2 r^3 \exp \left( -\pi \left( \kDI^2 \intF + \intM \right) r^2 
    \right).
\end{equation}
The success probability of a victim MUE is
\begin{equation}
  \begin{split}
    & \pr \left\{ \sir \!>\! \sir_0 \left| N_\text{DI}^{(\text{F})} \!=\! 1 \right. \right\} \\
    & = \ev_{I,\rM}^{(1)} \left\{ \pr \left\{ \frac{\pM h \rM^{-\pleM}}{I_\text{M}+\aRes \left( I_\text{DI} + I_\text{F} \right) } > \sir_0 \right\} \right\} \\
    & = \int_0^\infty \!\!\ev_I^{(1)} \left\{
    \pr \left\{ \frac{\pM h r^{-\pleM}}{I_\text{M} \!+\! \aRes \left( I_\text{DI} \!+\! I_\text{F} \right) } \!>\! \sir_0
    \right\} \right\} \\
    & \quad \times f_{\rM | \kDI}^{(1)}(r) dr.\label{eq_MF_absSuc}
  \end{split}
\end{equation}
The reader may notice that on RHS of \eqref{eq_MF_absSuc} we have omitted the condition $N_\text{DI}^{(\text{F})} \!=\! 1$. This is purely for space purposes and we will repeat it a few times throughout the paper. The inner probability from \eqref{eq_MF_absSuc} can be found as
\begin{align}
  & \ev_I^{(1)} \left\{ \right\} = \ev_I^{(1)} \left\{ \pr \left\{ \frac{\pM h
    r^{-\pleM}}{I_\text{M}+\aRes \left( I_\text{DI} + I_\text{F} \right) } > \sir_0 \right\} \right\} \\
  & = \ev_I^{(1)} \left\{ \pr \left\{ h > \frac{\sir_0 \left( I_\text{M}+\aRes
    \left( I_\text{DI} + I_\text{F} \right) \right) }{\pM r^{-\pleM}} \right\} \right\} \\
  & = \ev_I^{(1)} \left\{ \exp \left( -\frac{\sir_0 r^{\pleM}}{\pM} \left( I_\text{M}+\aRes
    I_\text{DI} + \aRes I_\text{F} \right) \right) \right\} \\
  & = \ev_{I_\text{M}}^{(1)} \left\{ \exp \left( -\frac{\sir_0 r^{\pleM}}{\pM} I_\text{M} \right) \right\}
  \ev_{I_\text{DI}}^{(1)} \left\{ \exp \left( -\frac{\sir_0 r^{\pleM}}{\pM} \aRes I_\text{DI} \right) \right\}
    \nonumber \\
  & \quad \times \ev_{I_\text{F}}^{(1)} \left\{ \exp \left( -\frac{\sir_0 r^{\pleM}}{\pM} \aRes I_\text{F} \right) \right\}
    \label{eq1}.
\end{align}
The first and the last terms in \eqref{eq1} have been derived in \cite{AnBaGa2011} and are given by
\begin{equation}
  \label{eq_I_M}
  \ev_{I_\text{M}}^{(1)} \left\{ \right\} =
    \exp \left( -\pi \intM r^2 \rho \left( \sir_0,\pleM \right) \right)
\end{equation}
and
\begin{equation}
  \label{eq_I_F_}
  \ev_{I_\text{F}}^{(1)} \left\{ \right\} =
    \exp \left( -\pi \kDI^2 \intF r^2 \rho \left( \frac{\sir_0 \aRes \pF r^{\pleM}}{\kDI^{\pleF} \pM r^{\pleF}} , \pleF \right) \right),
\end{equation}
where
\begin{equation}
  \rho \left( \gamma,\ple \right) = \int_{\gamma^\frac{-2}{\ple}}^\infty \frac{\gamma^\frac{2}{\ple}}{1+u^\frac{\ple}{2}}du.
\end{equation}
For averaging the dominant interference term in \eqref{eq1} we exploit the Laplace transform of exponential function as given here
\begin{align}
  & \ev_{I_\text{DI}}^{(1)} \left\{ \exp \left( -s I_\text{DI} \right) \right\} \nonumber \\
  & = \ev_{I_\text{DI}}^{(1)} \left\{ \exp \left( -s \pF h_\text{DI} r_\text{DI}^{-\pleF} \right) \right\} \\
  & = \ev_{r_\text{DI}}^{(1)} \left\{ \frac{1}{1+s \pF r_\text{DI}^{-\pleF}} \right\} \\
  & = \int_{r_\text{DI}} \frac{1}{1+s \pF u^{-\pleF}} f_{r_\text{DI}}(u) du \\
  & = \int_0^{\kDI r} \frac{1}{1+s \pF u^{-\pleF}} \frac{2 u}{\left( \kDI r \right)^2} du \\
  & = 1 - {}_2 F_1 \left( 1, \frac{2}{\pleF}, \frac{2 \!+\! \pleF}{\pleF}, -\frac{\left( \kDI r \right)^{\pleF}}{s \pF} \right),
\end{align}
where $h_\text{DI}$ and $r_\text{DI}$ are are fast fading power and link distance between victim MUE and DI FBS, respectively, and ${}_2 F_1 \left( \right)$ is the hypergeometric function. For $s \!=\! \frac{\sir_0 r^{\pleM} \aRes}{\pM}$ this gives
\begin{multline}
  \label{eq_I_1}
  \ev_{I_\text{DI}}^{(1)} \left\{ \exp \left( -\frac{\sir_0 r^{\pleM}}{\pM} \aRes I_\text{DI} \right) \right\} \\
    = 1 - {}_2 F_1 \left( 1, \frac{2}{\pleF}, \frac{2 \!+\! \pleF}{\pleF}, -\frac{\kDI^{\pleF} \pM r^{\pleF}}{\sir_0 \aRes \pF r^{\pleM}} \right).
\end{multline}
The PDF of $\rM$ \eqref{eq_rmCond} and interference terms \eqref{eq_I_M}, \eqref{eq_I_F_} and \eqref{eq_I_1} can now be plugged into \eqref{eq_MF_absSuc}. In case of single path loss exponent $\pleM \!=\! \pleF \!=\! \ple$ the integration variable disappears from inside the $\rho \left( \right)$ and ${}_2 F_1 \left( \right)$ functions and we get a closed form solution
\begin{multline}
  \label{eq_suc_femto_oneDI}
  \pr \left\{ \sir \!>\! \sir_0 \left| N_\text{DI}^{(\text{F})} \!=\! 1,\pleM \!=\! \pleF \!=\! \ple \right. \right\} \\
    = \frac{ \left( \kDI^2 \intF + \intM \right)^2
    \left( 1 - {}_2 F_1 \left( 1, \frac{2}{\ple}, \frac{2+\ple}{\ple}, -\frac{\kDI^\ple \pM}{\sir_0 \aRes \pF} \right)
    \right) }
    { \left( \intM \left[ 1+\rho \left( \sir_0,\ple \right) \right] +
    \kDI^2 \intF \left[ 1+\rho \left( \frac{\sir_0 \aRes \pF}{\kDI^\ple \pM},\ple \right) \right] \right)^2 }.
\end{multline}

\subsubsection{One or more dominant interferers}

A probability that an MUE has one or more FBS dominant interferers within $\kDI \rM$ distance is complementary to probability that there are no FBSs within that distance. Using the same approach as with the single DI we get the probability
\begin{equation}
  \label{eq_pr_onePlusDI}
  \pr \left\{ N_\text{DI}^{(\text{F})} \geq 1 \right\} = \frac{\kDI^2 \intF}{\kDI^2 \intF + \intM}
\end{equation}
and the PDF of distance from closest MBS
\begin{multline}
  f_{\rM | \kDI}(r) = \frac{\kDI^2 \intF \!+\! \intM}{\kDI^2 \intF} 2 \pi \intM r \\
    \times \bigg{(} \exp\left( -\pi \intM r^2 \right) \!-\! \exp\left( -\pi \left( \kDI^2 \intF \!+\! \intM \right) r^2 \right) \bigg{)}.
\end{multline}
The success probability is again given by
\begin{multline}
  \label{eq_suc_femto_onePlus}
  \pr \left\{ \sir \!>\! \sir_0 \left| N_\text{DI}^{(\text{F})} \geq 1 \right. \right\} \\
  = \int_0^\infty \!\!\ev_I \left\{
    \pr \left\{ \frac{\pM h r^{-\pleM}}{I_\text{M} \!+\! \aRes \left( I_\text{DI} \!+\! I_\text{F} \right) } \!>\! \sir_0
    \right\} \right\} f_{\rM | \kDI}(r) dr
\end{multline}
and the inner probability by
\begin{align}
  \ev_I \left\{ \right\} & = \ev_{I_\text{M}} \left\{ \exp \left( -\frac{\sir_0 r^{\pleM}}{\pM} I_\text{M} \right) \right\}
    \nonumber \\
  & \quad \times \ev_{I_\text{DI}} \left\{ \exp \left( -\frac{\sir_0 r^{\pleM}}{\pM} \aRes I_\text{DI} \right) \right\} \\
  & \quad \times \ev_{I_\text{F}} \left\{ \exp \left( -\frac{\sir_0 r^{\pleM}}{\pM} \aRes I_\text{F} \right) \right\}. \nonumber
\end{align}
The $I_\text{M}$ and $I_\text{F}$ terms generalized for arbitrary BS load values are given here:
\begin{equation}
  \ev_{I_\text{M}} \left\{ \right\} =
    \exp \left( -\pi \lM \intM r^2 \rho \left( \sir_0,\pleM \right) \right)
\end{equation}
\begin{equation}
  \ev_{I_\text{F}} \left\{ \right\} =
    \exp \left( -\pi \lF \kDI^2 \intF r^2 \rho \left( \frac{\sir_0 \aRes \pF r^{\pleM}}{\kDI^{\pleF} \pM r^{\pleF}} , \pleF \right) \right)
\end{equation}
To derive $\ev_{I_\text{DI}} \{ \}$ we will use $\ev_{I_\text{DI}}^{(1)} \{ \}$ from \eqref{eq_I_1}. We denote $I_\text{DI} \!=\! \sum_{i=1}^{N_\text{DI}^{(\text{F})}} I_\text{DI}^{(i)}$, where $N_\text{DI}^{(\text{F})}$ is a random variable describing the number of FBS DIs within $\kDI r$ conditioned on presence of at least one. With full load we can get an exact expression
\begin{align}
  \ev_{I_\text{DI}} \{ \} &= \ev \left\{ \left. \prod_{i=1}^{N_{DI}^{(F)}} \ev \left[ \exp \left( -\frac{\sir_0 r^{\pleM}}{\pM} \aRes I_\text{DI}^{(i)} \right) \right]
    \right| N_{DI}^{(F)} \right\} \\
  &= \ev \left\{ \ev_{I_\text{DI}}^{(1)} \{ \} ^{N_{DI}^{(F)}} \right\} \\
  &= \text{PGF}_{N_{DI}^{(F)}} \left( \ev_{I_\text{DI}}^{(1)} \{ \} \right),
\end{align}
where $\text{PGF}_{N_{DI}^{(F)}}$ is the probability generation function of $N_\text{DI}^{(\text{F})}$ given by
\begin{align}
  \text{PGF}_{N_{DI}^{(F)}} (x) &= \sum_{i=1}^{\infty} x^i f_{N_{DI}^{(F)}} (i) \\
  &= \sum_{i = 1}^\infty x^i \frac{\frac{(\pi \kDI ^2 \intF r^2)^i}{i!} \exp (-\pi \kDI ^2 \intF r^2)}{1 - \exp (-\pi \kDI ^2 \intF r^2)} \\
  &= \frac{\exp (x \pi \kDI ^2 \intF r^2) - 1}{\exp (\pi \kDI ^2 \intF r^2) - 1}
\end{align}
where $f_{N_{DI}^{(F)}} (n)$ is the PMF of $N_\text{DI}^{(\text{F})}$. With a general load $\lF$, the expression does not hold. In that case we can use a good approximation
\begin{multline}
  \ev_{I_\text{DI}} \left\{ \right\} \approx \ev_{I_\text{DI}}^{(1)} \left\{ \right\} ^{\lF \overline{N}_\text{DI}^{(\text{F})}} \label{eq_expIdi} \\
  = \left[ 1 - {}_2 F_1 \left( 1, \frac{2}{\pleF}, \frac{2 \!+ \! \pleF}{\pleF}, -\frac{\kDI^\pleF \pM r^{\pleF}}{\sir_0 \aRes \pF r^{\pleM}}
    \right) \right] ^{\lF \overline{N}_\text{DI}^{(\text{F})}},
\end{multline}
where $\overline{N}_\text{DI}^{(\text{F})}$ is the mean value of $N_\text{DI}^{(\text{F})}$ given by
\begin{equation}
  \overline{N}_\text{DI}^{(\text{F})} = \frac{\pi \kDI^2 \intF r^2}{1-\exp \left( -\pi \kDI^2 \intF r^2 \right)}.
\end{equation}
The interference terms can now be put into \eqref{eq_suc_femto_onePlus} to calculate the success probability. Unlike \eqref{eq_suc_femto_oneDI}, the integral in \eqref{eq_suc_femto_onePlus} cannot be simplified into a more digestible form even with single path loss exponent and has to be evaluated numerically.

\subsection{Macro/pico scenario}

In the second scenario, the downlink signal-to-interference-ratio at a victim PUE is defined as
\begin{equation}
  \sir = \frac{\pP h \rP^{-\pleP}}{I_\text{P} + \aRes \left( I_\text{DI} + I_\text{M} \right)},
\end{equation}
where $I_\text{P}$ is sum interference power from the pico tier (all PBSs except associated one), $I_\text{DI}$ is sum interference power from dominant MBS interferers and $I_\text{M}$ is sum interference power from all other MBS interferers. In the rest of the subsection we will state the most important results for the success probability. Derivation follows the same logic as in the first scenario.

The probability that a UE is actually a victim PUE is
\begin{equation}
  \label{eq_pr_vicPUE}
  \pr \left\{ k_2 \rM < \rP < k_1 \rM \right\} = \frac{\intM}{\intM + \kDI_2^2 \intP} - \frac{\intM}{\intM + \kDI_1^2 \intP}
\end{equation}
and the PDF of $\rP$ of a victim PUE
\begin{multline}
  f_{\rP | \kDI}(r) = 2 \pi r \frac{\left( \intM + \kDI_1^2 \intP \right)  \left( \intM + \kDI_2^2 \intP \right)}
    {\left( \kDI_1^2 - \kDI_2^2 \right) \intM} \\
    \times \left[ \exp \left( -\pi \left( \frac{\intM}{\kDI_1^2} \!+\! \intP \right) r^2 \right) \!-\!
    \exp \left( -\pi \left( \frac{\intM}{\kDI_2^2} \!+\! \intP \right) r^2 \right) \right].
\end{multline}
The success probability is given by integral
\begin{equation}\label{eq_macroPicoSuccess}
  \begin{split}
  & \pr \left\{ \sir \!>\! \sir_0 \left| k_2 \rM \!<\! \rP \!<\! k_1 \rM \right. \right\} \\
  & = \int_0^\infty \!\!\ev_I \left\{
    \pr \left\{ \frac{\pP h r^{-\pleP}}{I_\text{P} \!+\! \aRes \left( I_\text{DI} \!+\! I_\text{M} \right) } \!>\! \sir_0
    \right\} \right\} \\
  & \quad \times f_{\rP | \kDI}(r) dr
  \end{split}
\end{equation}
with the inner probability
\begin{align}
  \ev_I \left\{ \right\} & = \ev_{I_\text{P}} \left\{ \exp \left( -\frac{\sir_0 r^{\pleP}}{\pP} I_\text{P} \right) \right\}
    \nonumber \\
  & \quad \times \ev_{I_\text{DI}} \left\{ \exp \left( -\frac{\sir_0 r^{\pleP}}{\pP} \aRes I_\text{DI} \right) \right\} \\
  & \quad \times \ev_{I_\text{M}} \left\{ \exp \left( -\frac{\sir_0 r^{\pleP}}{\pP} \aRes I_\text{M} \right) \right\}. \nonumber
\end{align}
The non-dominant interference terms $\ev_{I_\text{P}}$ and $\ev_{I_\text{M}}$ are given here:
\begin{equation}
  \ev_{I_\text{P}} \left\{ \right\} =
    \exp \left( - \pi \lP \intP r^2 \rho \left( \sir_0 , \pleP \right) \right)
\end{equation}
\begin{equation}
  \ev_{I_\text{M}} \left\{ \right\} =
    \exp \left( - \pi \lM \frac{\intM}{\kDI_2^2} r^2 \rho \left( \frac{\sir_0 \kDI_2^{\pleM} \aRes \pM r^{\pleP}}{\pP r^{\pleM}},\pleM \right) \right)
\end{equation}
Fully loaded dominant interference term can by calculated using
\begin{equation}
  \ev_{I_\text{DI}} \left\{ \right\} = \frac{\exp \left( \ev_{I_\text{DI}}^{(1)} \{ \} \pi \left( \frac{1}{\kDI _2^2} - \frac{1}{\kDI _1^2} \right) \intM r^2 \right) - 1}
    {\exp \left( \pi \left( \frac{1}{\kDI _2^2} - \frac{1}{\kDI _1^2} \right) \intM r^2 \right) - 1}
\end{equation}
where
\begin{equation}
  \begin{split}
  & \ev_{I_\text{DI}}^{(1)} \left\{ \right\} = \Bigg{(} \frac{\kDI_1 \kDI_2}{\kDI_1^2 - \kDI_2^2} \\
  & \times \left[ \frac{1}{\kDI_2^2}
    {}_2 F_1 \left( 1,-\frac{2}{\pleM},\frac{\pleM \!-\! 2}{\pleM},\frac{-\sir_0 \kDI_2^{\pleM} \aRes \pM r^{\pleP}}{\pP r^{\pleM}} \right) 
      \right. \\
  & \left. {}-{} \frac{1}{\kDI_1^2} {}_2 F_1 \left( 1,-\frac{2}{\pleM},\frac{\pleM \!-\! 2}{\pleM},\frac{-\sir_0 \kDI_1^{\pleM} \aRes \pM
      r^{\pleP}}{\pP r^{\pleM}} \right) \right] \Bigg{)}. 
  \end{split}
\end{equation}
For a general load $\lM$ we then have
\begin{equation}
  \ev_{I_\text{DI}} \left\{ \right\} \approx \ev_{I_\text{DI}}^{(1)} \left\{ \right\} ^{\lM \overline{N}_\text{DI}^{(\text{M})}},
\end{equation}
where $\overline{N}_\text{DI}^{(\text{M})}$ is the average number of MBS DIs calculated as
\begin{equation}
  \overline{N}_\text{DI}^{(\text{M})} = \frac{\pi \intM \left( \frac{1}{\kDI_2^2}-\frac{1}{\kDI_1^2} \right) r^2}
    {1-\exp \left( - \pi \intM \left( \frac{1}{\kDI_2^2}-\frac{1}{\kDI_1^2} \right) r^2 \right)}
\end{equation}
Like in the first scenario, even with single path loss exponent the final integral \eqref{eq_macroPicoSuccess} can be evaluated only numerically.

\section{Number of ABSFs}
\label{sec:numAbsfs}

3GPP is introducing ABSFs to protect victim UEs. In both considered scenarios, victim UEs as we defined them are easily identified by presence of dominant interferers in the vicinity based on RSRP measurements. We can thus focus all our attention on the victim UEs and avoid a general and complicated sum rate optimization problem with a simplified one
\begin{align}
  \underset{\nAbsf}{\text{maximize}} &~~ C_\text{NV} \left( \nAbsf \right) \nonumber \\
  \text{subject to} &~~ C_\text{V} \left( \nAbsf \right) \geq C_\text{V,min}, \label{eq_opt1}\\
  &~~ 0 \leq \nAbsf \leq \nSf, \nonumber
\end{align}
where $\nSf$ represents number of subframes in a frame, $\nAbsf$ represents number of ABSFs in a frame, $C_\text{NV}$ stands for throughput of non-victim UEs, $C_\text{V}$ stands for victim UE throughput and $C_\text{V,min}$ stands for the minimum required victim UE throughput, a parameter of choice. We will not define $C_\text{NV} \left( \nAbsf \right)$ more closely because any reasonable definition is a strictly decreasing function of $\nAbsf$. We thus turn even more attention to the victim UEs and define a greatly simplified problem
\begin{align}
  {\text{minimize}} &~~ \nAbsf \nonumber \\
  \text{subject to} &~~ C_\text{V} \left( \nAbsf \right) \geq C_\text{V,min}, \label{eq_opt2}\\
  &~~ 0 \leq \nAbsf \leq \nSf. \nonumber
\end{align}
Now, the throughput of victim UEs depends on quite many things, from which the most important one is how the BS schedules the UEs, i.e., what is the scheduling algorithm and how are victim and non-victim UE transmissions placed in NSFs and ABSFs. In \cite{WaPe2011} it has been suggested that proportionally fair scheduler, assuming reliable knowledge of channel state, can take care of the division between NSFs and ABSFs itself. However, scheduling is (and will probably long stay) implementation specific, so that BS hardware and software vendors can compete with each other and fight for the technological edge.

Therefore, to define $C_\text{V} \left( \nAbsf \right)$, we consider what is from the throughput and scheduling point of view the worst case scenario: average throughput at outage threshold (outage throughput) and round robin scheduling algorithm. Concerning NSF and ABSF division we assume that non-victim UEs have access only to NSFs while victim UEs have access to both NSFs and ABSFs. With these the average victim UE outage throughput is
\begin{equation}
  C_\text{V} = \ev_{\nUE,\nUE_\text{V}} \left\{ \frac{\nAbsf}{\nSf} C_\text{V}^{(\text{A})} \!\left( \nUE,\nUE_\text{V} \right) +
    \frac{\nSf-\nAbsf}{\nSf} C_\text{V}^{(\text{N})} \!\left( \nUE,\nUE_\text{V} \right) \right\}
\end{equation}
where $C_\text{V}^{(\text{A})}$ is average victim UE outage throughput in ABSFs, $C_\text{V}^{(\text{N})}$ is average victim UE outage throughput in NSFs and $\nUE$ and $\nUE_\text{V}$ represent the number of all UEs and the number of victim UEs associated with the BS, respectively. Although $\nUE$ and $\nUE_\text{V}$ are correlated, for our purposes the outage throughput $C_\text{V}$ can be very well approximated by considering them as independent. We thus write
\begin{equation}
  C_\text{V} \approx \frac{\nAbsf}{\nSf} C_\text{V}^{(\text{A})} \!\left( \nUE_\text{V} \right) + \frac{\nSf-\nAbsf}{\nSf}
    C_\text{V}^{(\text{N})} \!\left( \nUE \right).\label{eq_oTp}
\end{equation}
Now, assuming channel independence across resource blocks and using the success probabilities derived in Section \ref{sec_succProb}, we can approximate $C_\text{V}^{(\text{A})}$ and $C_\text{V}^{(\text{N})}$ as
\begin{equation}
  C_\text{V}^{(\text{A})} \!\left( \nUE_\text{V} \right) \approx \nRb \pr \left\{ \sir^{(\text{A})} > \sir_0 \right\} \log \left( 1+\sir_0 \right) 
    \overline{\Omega}^{(\text{A})},
\end{equation}
\begin{equation}
  C_\text{V}^{(\text{N})} \!\left( \nUE \right) \approx \nRb \pr \left\{ \sir^{(\text{N})} > \sir_0 \right\} \log \left( 1+\sir_0 \right)
    \overline{\Omega}^{(\text{N})},
\end{equation}
where $\nRb$ is number of resource blocks, the $\overline{\Omega}^{(\text{A})}$ and $\overline{\Omega}^{(\text{N})}$ terms denote the average asymptotic proportion of resources that a victim UE is scheduled via the round robin principle in ABSF and NSF, respectively, and $\sir^{(\text{A})}$ and $\sir^{(\text{N})}$ denote SIR in ABSF and NSF, respectively. In place of the success probabilities we use the corresponding macro/femto and macro/pico derived terms \eqref{eq_suc_femto_onePlus} and \eqref{eq_macroPicoSuccess}, respectively. The $\overline{\Omega}^{(\text{A})}$ and $\overline{\Omega}^{(\text{N})}$ values depend on the average number of victim and non-victim UEs per BS. Derivation of these values is presented in the Appendix. Finally, we connect all the acquired results and construct a condition for the number of ABSFs as
\begin{equation}
  \label{eq_numAbsf}
  \nAbsf = \min \left[ \left\lceil \frac{\nSf \left( C_\text{V,min}-C_\text{V}^{(\text{N})} \right) }{C_\text{V}^{(\text{A})}
    - C_\text{V}^{(\text{N})}} \right\rceil ~;~ \nSf \right].
\end{equation}

\subsection{Macro/femto scenario}
\label{ssec:numAbsfMF}

\begin{table}[t!]
  \caption{Reference simulation parameters}
  \label{tab_params}
  \begin{center}
      \begin{tabular}{cc}
        \hline
        Parameter & Value \\
	    \hline
	    \hline
		Simulated area & $10000\text{m} \times 10000\text{m}$ \\
		Samples collected from & $3000\text{m} \times 3000\text{m}$ \\
		Macro BS intensity $\intM$ & $10^{-5}\text{m}^{-2}$ \\
		Femto BS intensity $\intF$ & $12\intM$ \\
		Pico BS intensity $\intP$ & $4\intM$ \\
		Macro UE intensity $\intMUE$ & $20\intM$ \\
		Macro BS transmission power $\pM$ & 43dBm \\
		Femto BS transmission power $\pF$ & 20dBm \\
		Pico BS transmission power $\pP$ & 30dBm \\
		Macro BS load $\lM$ & 1 \\
		Femto BS load $\lF$ & 0.5 \\
		Pico BS load $\lP$ & 0.8 \\
		MBS-UE path loss exponent $\pleM$ & 2.5 \\
		FBS-UE path loss exponent $\pleF$ & 3.5 \\
		PBS-UE path loss exponent $\pleP$ & 3 \\
		Macro/femto DI-defining $\kDI$ & $\left( \frac{\pF}{\pM} \right)^{\frac{2}{\pleM+\pleF}} \!=\! 0.136$ \\
		Macro/pico association bias $\cae$ & 7dB \\
		Macro/pico association-defining $\kDI_1$ & $\left( \frac{\cae \pP}{\pM} \right)^{\frac{2}{\pleM+\pleP}} \!=\! 0.471$ \\
		Macro/pico DI-defining $\kDI_2$ & $\left( \frac{\pP}{\pM} \right)^{\frac{2}{\pleM+\pleP}} \!=\! 0.262$ \\
		ABSF residue interference $\aRes$ & $-20\text{dB}$ \\
		Outage threshold $\sir_0$ & $-5\text{dB}$ \\
		Number of subframes $\nSf$ & 10 \\
		Number of resource blocks $\nRb$ & 25 \\
		Resource block bandwidth & $180\text{kHz}$ \\
		\hline
	  \end{tabular}
  \end{center}
\end{table}  

\begin{figure}[t!]
  \begin{center}
    \includegraphics[scale=0.5]{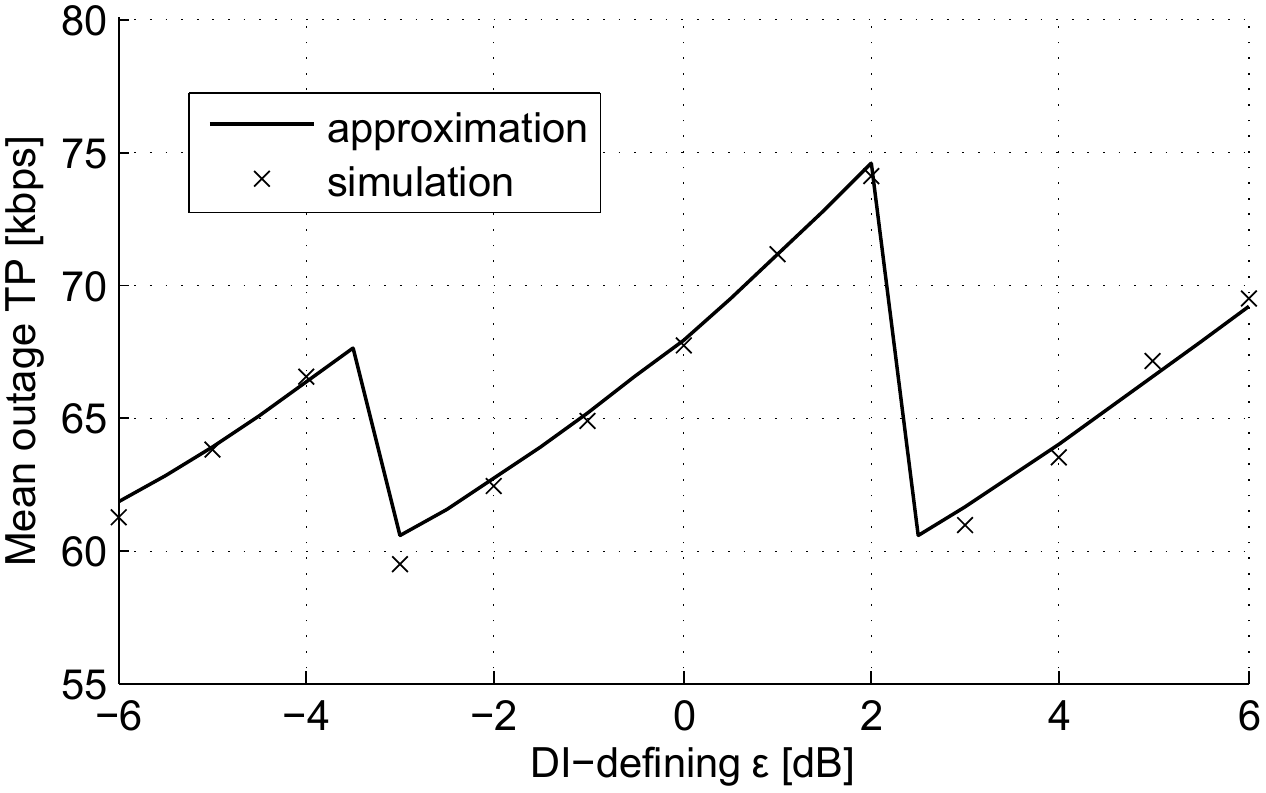}
    \caption{Approximate outage throughput of victim MUEs in macro/femto scenario versus DI definition via $\mf$ (as in $\kDI \!=\! \left[ \pF / \left( \mf \pM \right) \right]^{2 / \left( \pleM \!+\! \pleF \right)}$) compared to simulated equivalent.}\label{fig_apprx_oTp}
  \end{center}
\end{figure}

Before moving on to performance evaluation, we demonstrate precision of our outage throughput approximation and analyze dependence of the condition \eqref{eq_numAbsf} on key parameters. Unless stated otherwise, parameters are at their reference values as shown in Table \ref{tab_params}. We consider these to be realistic values. Because $\pleM \!\neq\! \pleF$ and $\pleM \!\neq\! \pleP$, we decided to use $\left( \pleM \!+\! \pleF \right) / 2$ and $\left( \pleM \!+\! \pleP \right) / 2$ in the definitions of $\kDI$, $\kDI_1$ and $\kDI_2$. That way we keep the definitions close to the intuitive shape with single $\ple$. Impact of different DI-defining coefficients is later considered in the performance evaluation. Unless stated otherwise, the minimum average outage throughput $C_\text{V,min}$ in macro/femto scenario is set to 40kbits/s.

In Fig. \ref{fig_apprx_oTp} we plot the approximation of outage throughput compared to simulated equivalent. Macrocell path loss and minimum average outage throughput are changed from default values to $\pleM \!=\! 3$ and $C_\text{V,min} \!=\! 60\text{kbits/s}$ in order to increase simulation precision (for further discussion see Section \ref{sec_perf}). Other parameters are at their default values as in Table \ref{tab_params}. The values of $C_\text{V}$ are plotted against definition of DI. The horizontal axis does not present $\kDI$ directly, but a ratio $\mf$ that puts into relation own received power and DI received power, i.e., $\kDI \!=\! \left[ \pF / \left( \mf \pM \right) \right]^{2 / \left( \pleM \!+\! \pleF \right)}$. The approximation can be considered very good, with deviations coming mostly from insufficient number of Monte Carlo samples. Within $\mf \!\in\! \left( -4\text{dB},-3\text{dB} \right)$ and $\mf \!\in\! \left( 2\text{dB},3\text{dB} \right)$ there is a step in the curve because for given values of $\kDI$ the number of required ABSFs according to our condition \eqref{eq_numAbsf} changes. 

\begin{figure}[t!]
  \begin{center}
    \includegraphics[scale=0.35]{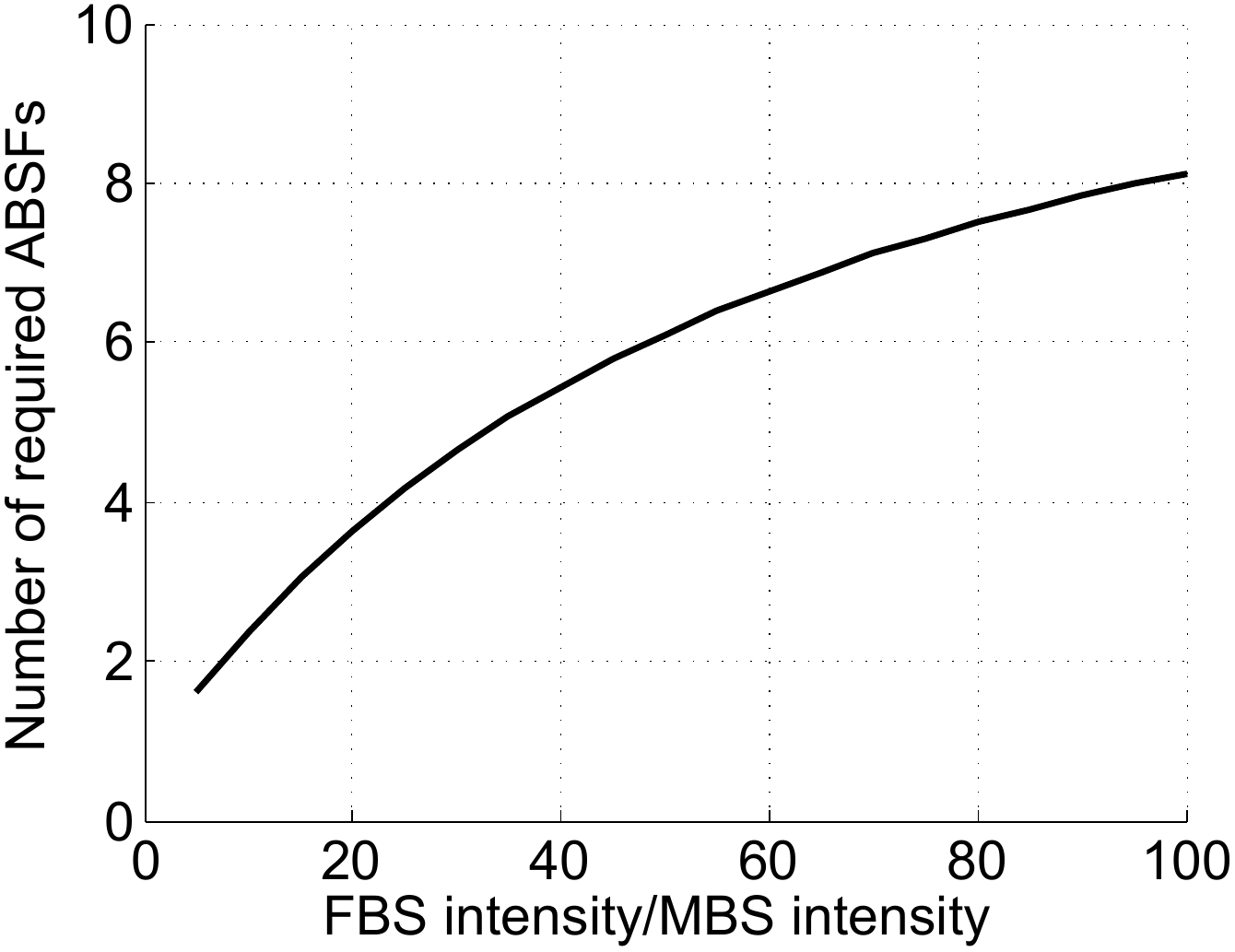}
    \caption{Dependence of required number of ABSFs $\nAbsf$ in macro/femto scenario on FBS intensity $\intF$ with other parameters constant. The $\intF / \intM$ ratio equals mean number of FBSs per MBS coverage.}\label{fig_sens_lambda_f}
  \end{center}
\end{figure}
\begin{figure}[t!]
  \begin{center}
    \includegraphics[scale=0.35]{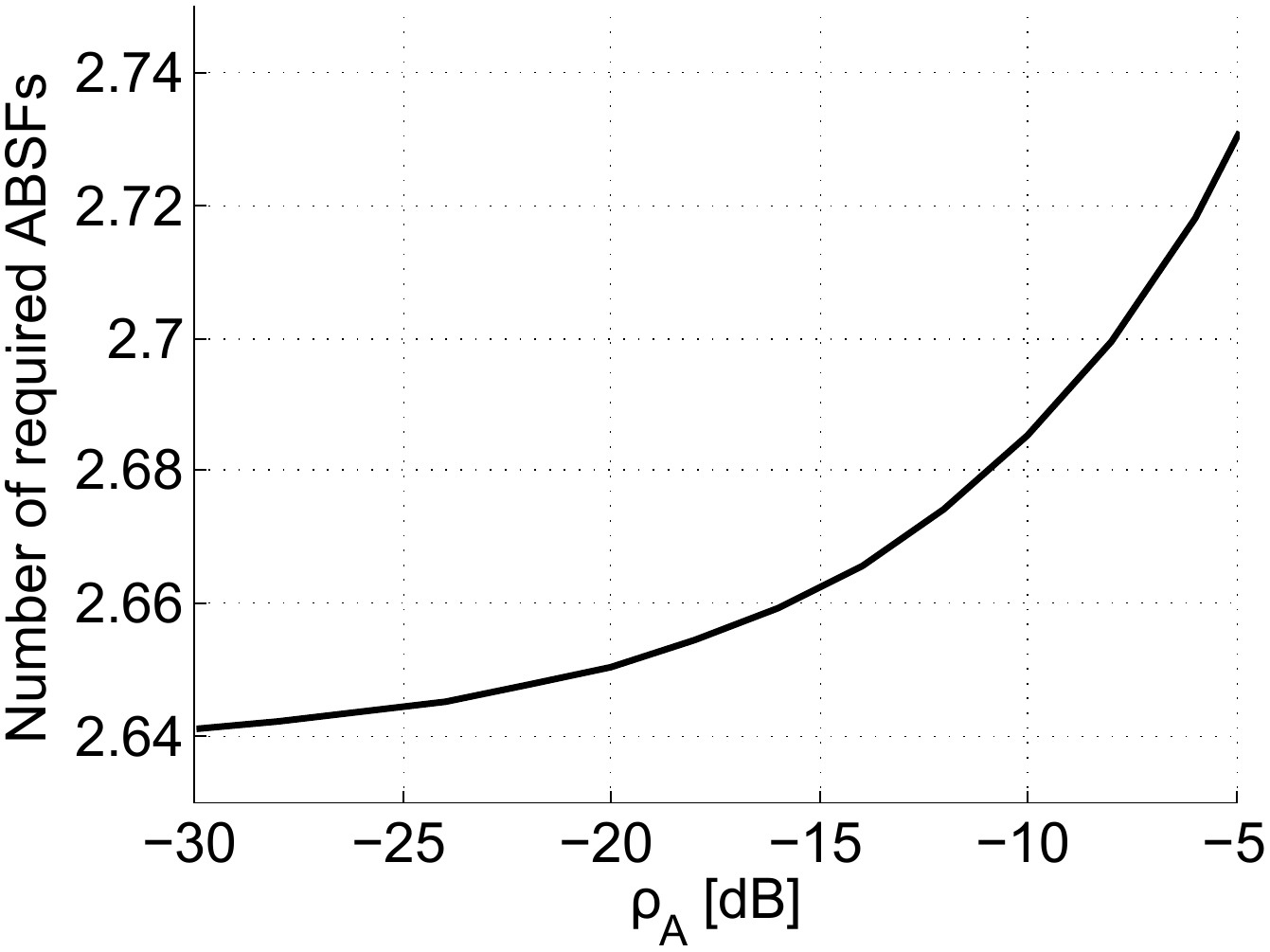}
    \caption{Dependence of the required number of ABSFs $\nAbsf$ in macro/femto scenario on residue ABSF interference $\aRes$ in macro/femto scenario with other parameters constant.}\label{fig_sens_rho_A}
  \end{center}
\end{figure}

\begin{figure}[t!]
  \begin{center}
    \includegraphics[scale=0.35]{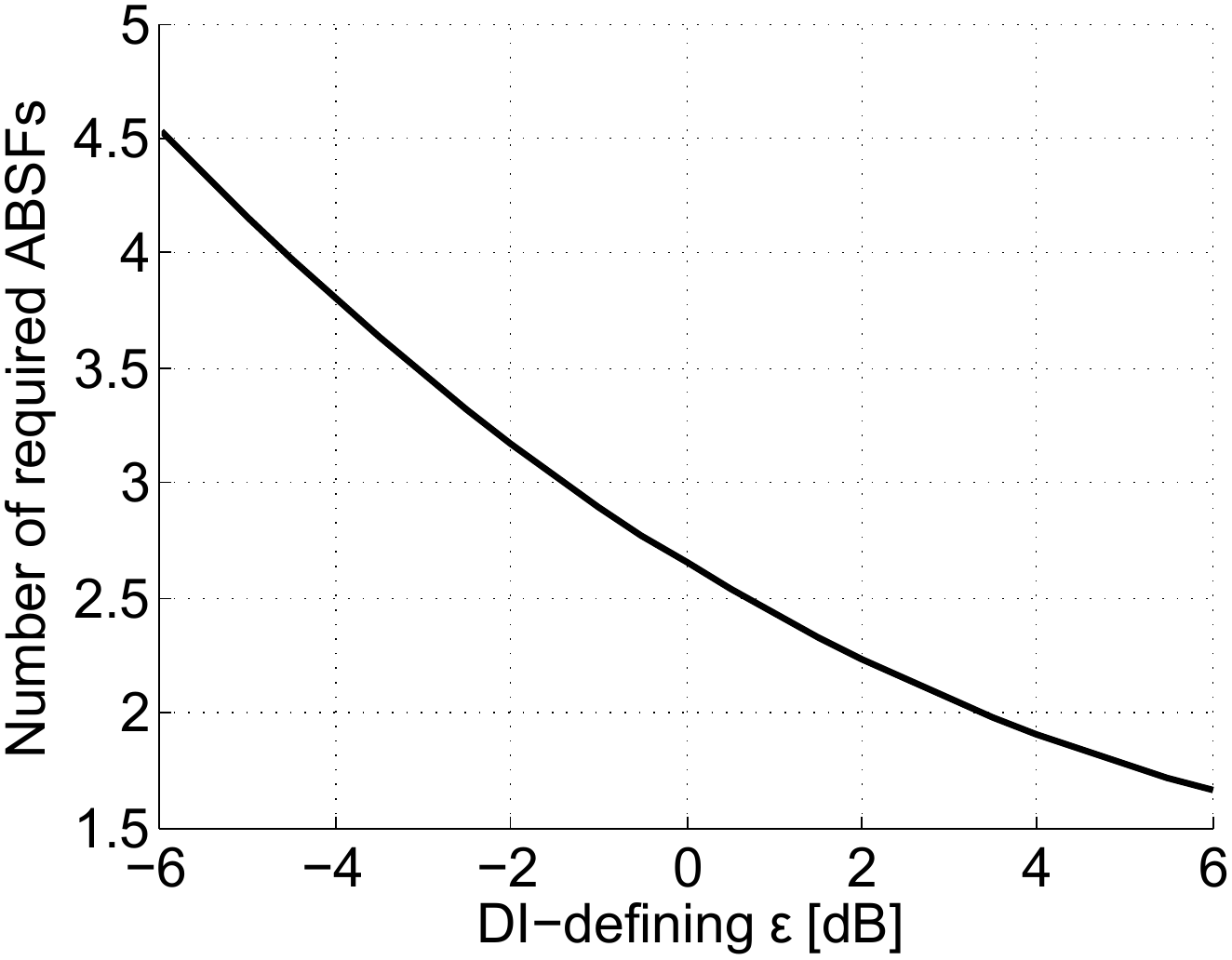}
    \caption{Dependence of the required number of ABSFs $\nAbsf$ in macro/femto scenario on the definition of victim MUE with other parameters constant.}\label{fig_sens_k}
  \end{center}
\end{figure}

Dependence of \eqref{eq_numAbsf} on $C_\text{V,min}$ and $\intMUE$ is intuitively clear and our results confirm it - the $\nAbsf$ required grows approximately linearly with both of these values. Therefore, we save space and exclude those results. In Fig. \ref{fig_sens_lambda_f} we show dependence on $\intF$. The trend is approximately linear in the beginning and then becomes slightly saturated. In Fig. \ref{fig_sens_rho_A} we show dependence on $\aRes$, the residual FBS interference in ABSF. Here the dependence is low because FBS interference is strongly attenuated by larger path loss exponent $\pleF$. Finally in Fig. \ref{fig_sens_k} we show dependence of $\nAbsf$ required on the definition of dominant FBS interferer $\kDI$ via $\mf$. With increasing $\mf$ the $\kDI$ coefficient is decreasing and less MUEs are considered being victim. Although this decreases the $\nAbsf$ required, we are practically increasing the number of non-victim MUEs and therefore have to consider effect on their performance. We will come back to this in Section \ref{sec_perf}.

\subsection{Macro/pico scenario}

\begin{figure}[t!]
  \begin{center}
    \includegraphics[scale=0.35]{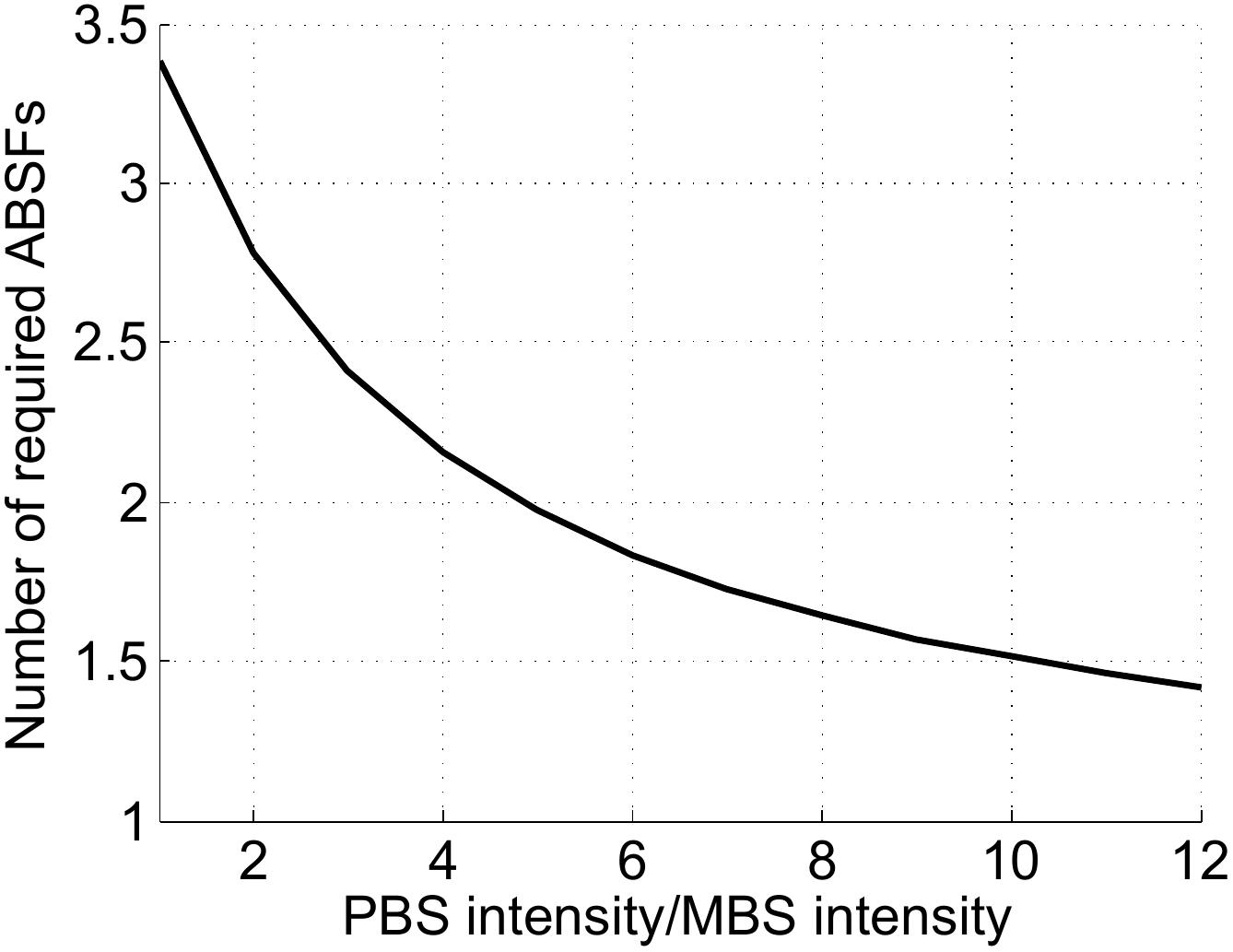}
    \caption{Dependence of the required number of ABSFs $\nAbsf$ in macro/pico scenario on PBS intensity $\intP$ with other parameters constant. The $\intP / \intM$ ratio equals mean number of PBSs per MBS coverage.}\label{fig_sens_lambda_p}
  \end{center}
\end{figure}

Similarly to the previous subsection, we present dependence of $\nAbsf$ on selected parameters. The reference parameter values are taken from Table \ref{tab_params} and the minimum average outage throughput $C_\text{V,min}$ is set to 100kbits/s.

\begin{figure}[t!]
  \begin{center}
    \includegraphics[scale=0.35]{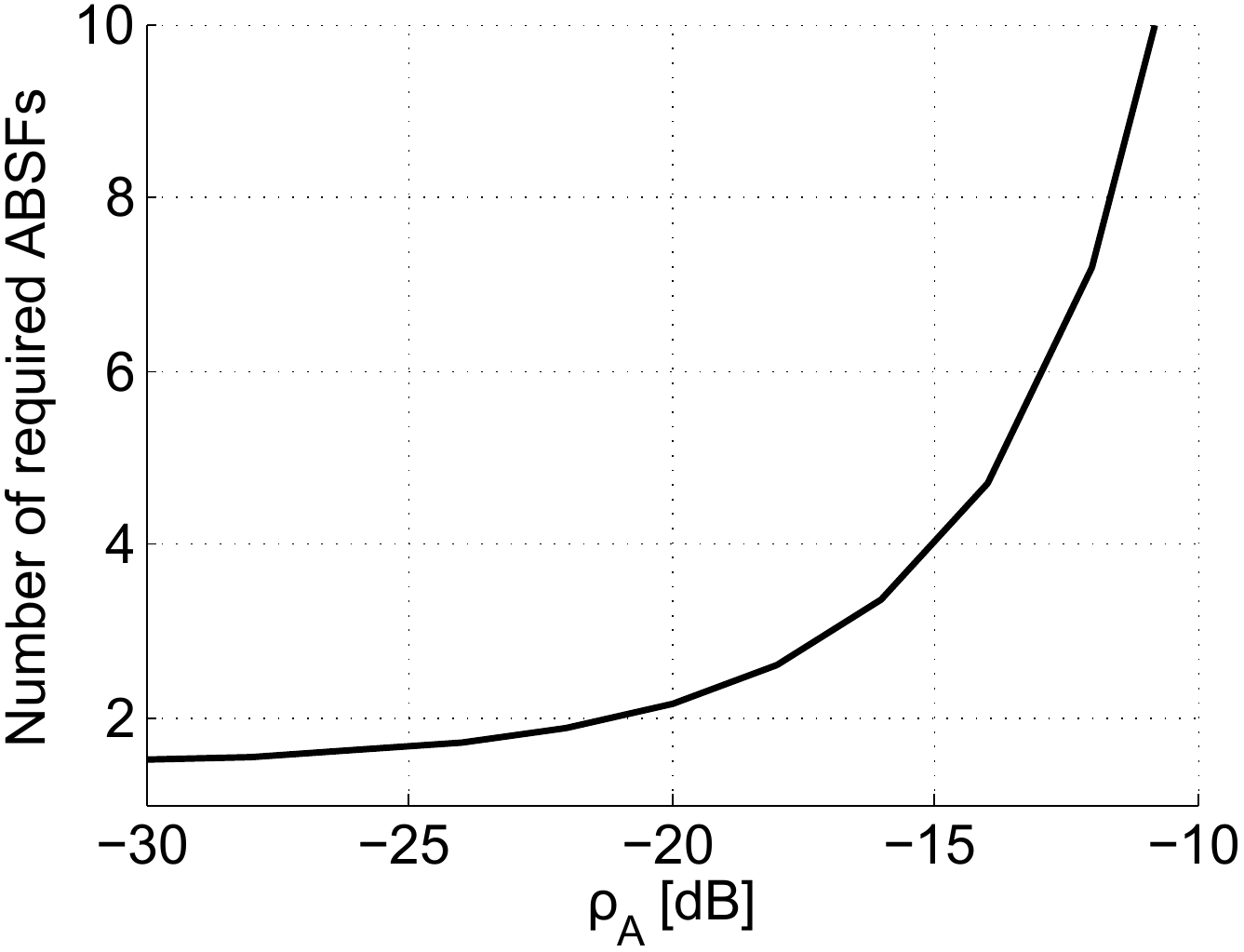}
    \caption{Dependence of the required number of ABSFs $\nAbsf$ in macro/pico scenario on residue ABSF interference $\aRes$ with other parameters constant.}\label{fig_sens_mp_rho_A}
  \end{center}
\end{figure}

\begin{figure}[t!]
  \begin{center}
    \includegraphics[scale=0.35]{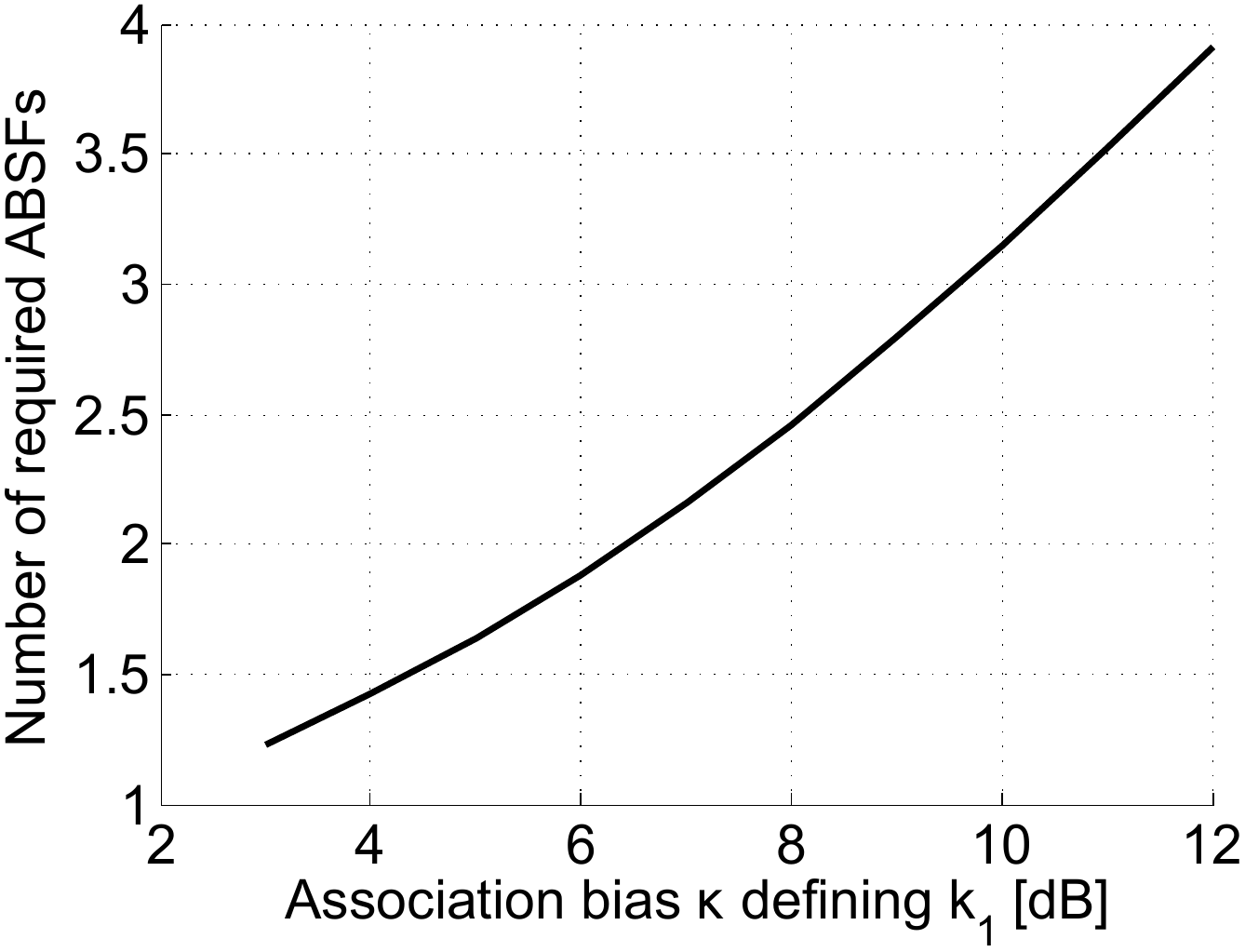}
    \caption{Dependence of the required number of ABSFs $\nAbsf$ in macro/pico scenario on association bias $\cae$, which defines $\kDI_1$, with other parameters constant.}\label{fig_sens_k_1}
  \end{center}
\end{figure}

In Fig. \ref{fig_sens_lambda_p} we show the effect of $\intP$ value on $\nAbsf$ with other parameters at reference values. Compared to intensity of FBSs in macro/femto scenario, increasing number of PBSs does not primarily increase interference. On the contrary, with higher number of PBSs the MUEs have more BSs to connect to, therefore number of PUEs per PBS decreases and less ABSFs is needed. In Fig. \ref{fig_sens_mp_rho_A} we show dependence of $\nAbsf$ on residue ABSF interference $\aRes$. While below $\aRes \!=\! -15\text{dB}$ the residue interference seems manageable, higher values leads to dramatic increase in the $\nAbsf$ requirement. The impact is much stronger than in macro/femto scenario. It is partly because of lower path loss exponent on MBS-UE links $\pleM \!<\! \pleF$ and partly because of the vulnerability imposed by the association bias $\cae$. Finally in Fig. \ref{fig_sens_k_1} we present dependence of $\nAbsf$ on $\cae$. As intuitively expected, increasing $\cae$ results in MBS interferers being even closer to the victim PUE and thus the $\nAbsf$ requirement increases.

\section{Performance analysis}
\label{sec_perf}

In this section we demonstrate the effect of the derived number of ABSFs on UE throughput in downlink. We use Monte Carlo simulations to evaluate UE throughput with link adaptation modeled by Shannon's capacity. The simulations consist of snapshots, during which BSs and UEs are dropped and kept at fixed positions, and in each snapshot there are multiple frames (with subframes) where the UEs are scheduled. Reference simulation parameters are summarized in Table \ref{tab_params}. Although we consider the model parameters realistic, the computational complexity of the simulations prevented us from using a simulated area size that would provide sufficient precision for $\pleM \!=\! 2.5$. However, rather than using unrealistic parameters we keep the area at manageable value and accept that the performance results in this section are a little on the optimistic side. As a measure of performance we collect aggregate throughput of each MUE/PUE and then evaluate the results in form of UE throughput CDF.

\subsection{Macro/femto scenario}

\begin{figure}[t!]
  \begin{center}
    \includegraphics[scale=0.5]{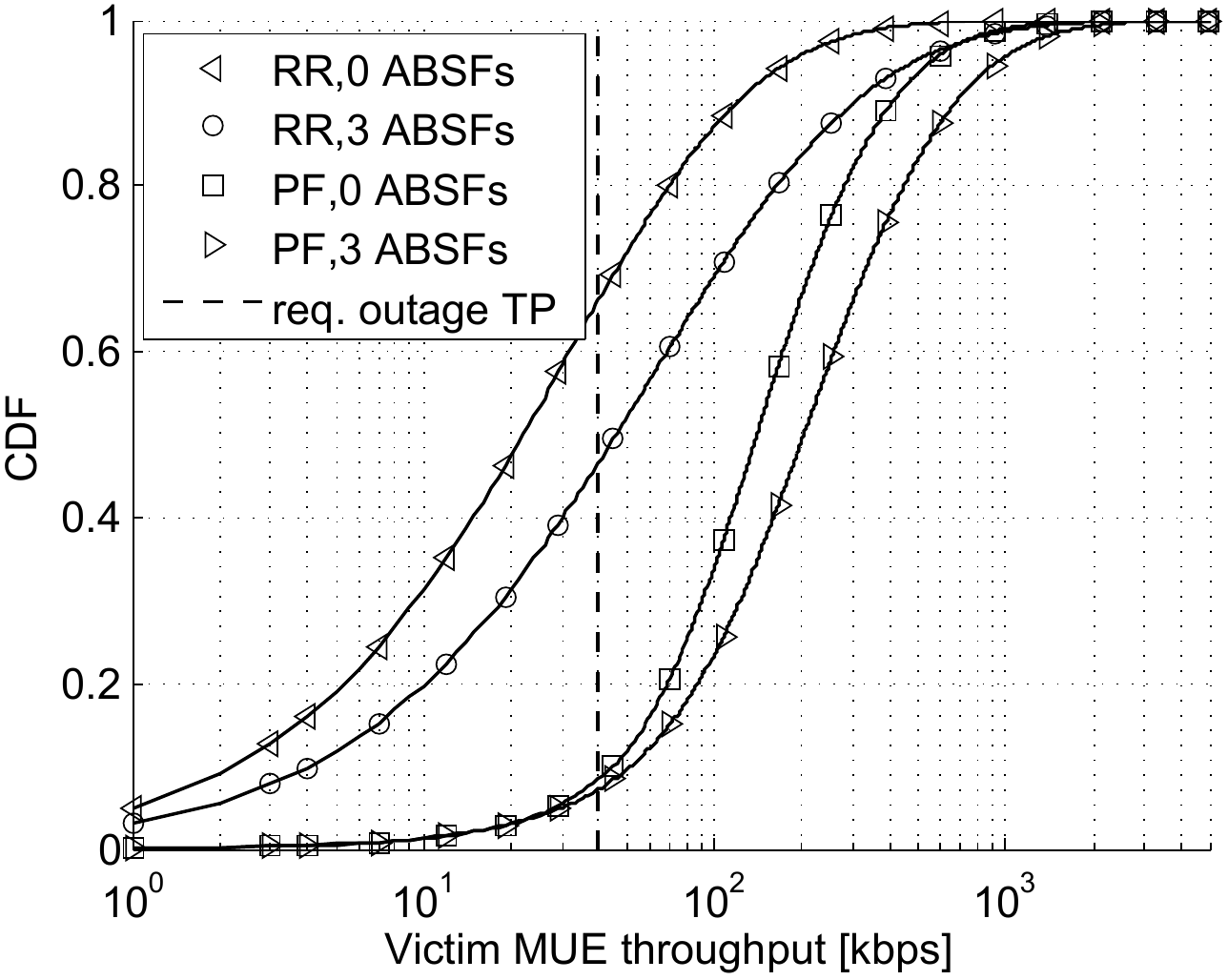}
    \caption{Throughput CDF of victim MUEs in macro/femto scenario with round robin and proportional fair scheduling, with and without ABSFs. The required outage throughput is marked by a dashed line.}\label{fig_mf_tpCdf_vic}
  \end{center}
\end{figure}
\begin{figure}[t!]
  \begin{center}
    \includegraphics[scale=0.5]{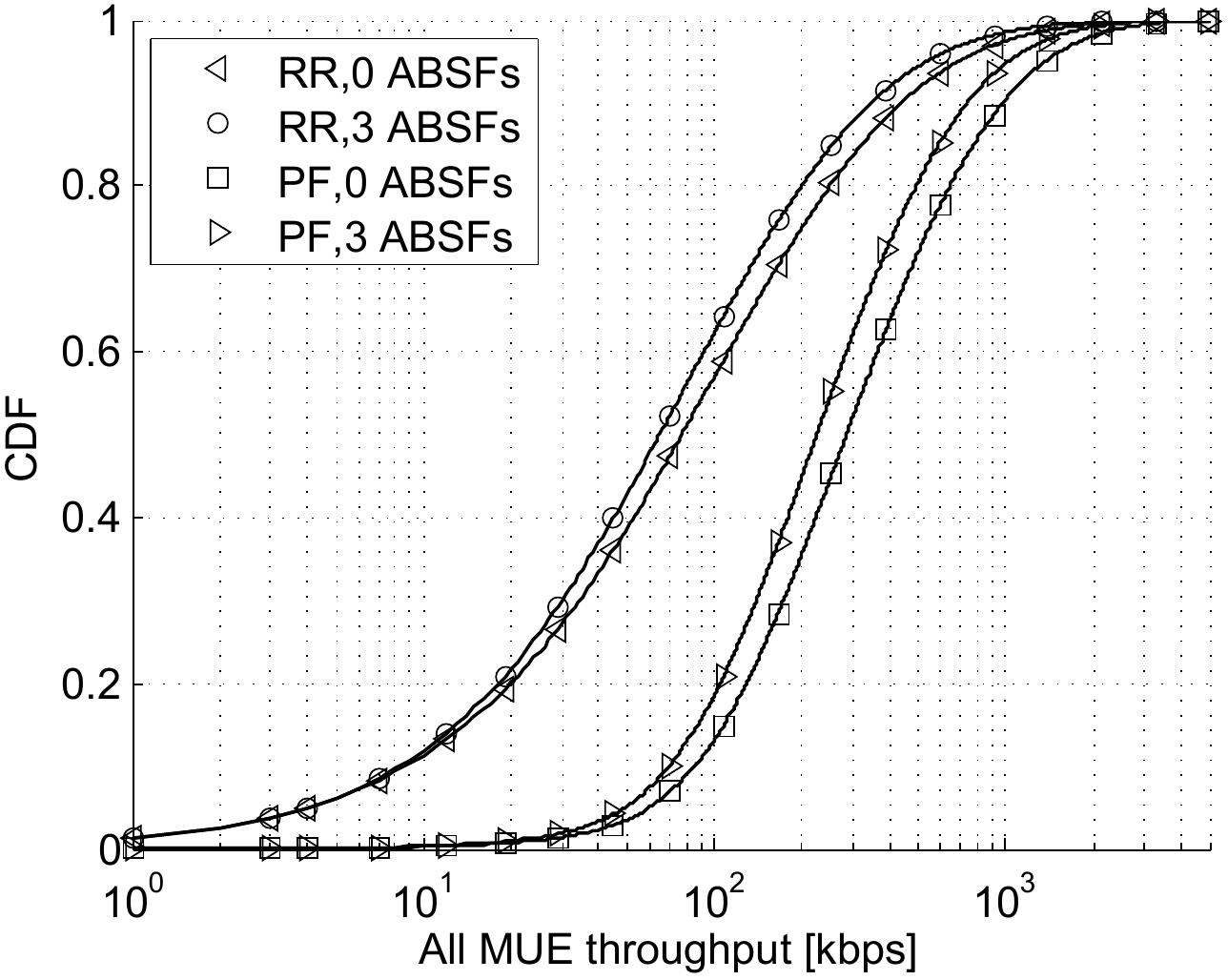}
    \caption{Throughput CDF of all MUEs in macro/femto scenario with round robin and proportional fair scheduling, with and without ABSFs.}\label{fig_mf_tpCdf_all}
  \end{center}
\end{figure}

The minimum average outage throughput in macro/femto scenario is set to $40\text{kbits/s}$. In Fig. \ref{fig_mf_tpCdf_vic} we present throughput CDFs of victim MUEs. Victim MUEs are scheduled in both NSFs and ABSFs and for comparison we have included both round robin and proportionally fair scheduling algorithms. The proportionally fair scheduler requires channel knowledge and in each RB it chooses the user with highest instantaneous SIR normalized by its mean SIR. The dashed line in Fig. \ref{fig_mf_tpCdf_vic} represents the chosen minimum mean victim outage throughput $C_\text{V,min}$.

Looking at the victim MUE curves we can observe that in given macro/femto scenario the effect of ABSFs is rather moderate. Without them, around 70\% of victim MUEs have lower throughput than $C_\text{V,min}$. With ABSFs, approximately 50\% have higher throughput than $C_\text{V,min}$. With proportionally fair scheduler the throughput values are higher and effect of ABSFs weaker. The large difference between our round robin based requirement and the proportionally fair performance suggests that our research should be expanded by considering advanced scheduling during the ABSF planning phase. The effect of ABSFs on purely non-victim MUEs is not shown, as those obviously lose 30\% of available resources. In Fig. \ref{fig_mf_tpCdf_all} we show throughput CDFs of all MUEs combined. The rather small improvement of victim MUE performance seems to be in overall statistics completely overshadowed by the effect of resource restrictions for non-victim users. From the system perspective these findings suggest that rather than blocking femto layer users and non-victim MUEs from a fraction of resources, the MUEs that suffer from strong FBS interference should be treated in an individual manner.

\begin{figure}[t!]
  \begin{center}
    \includegraphics[scale=0.5]{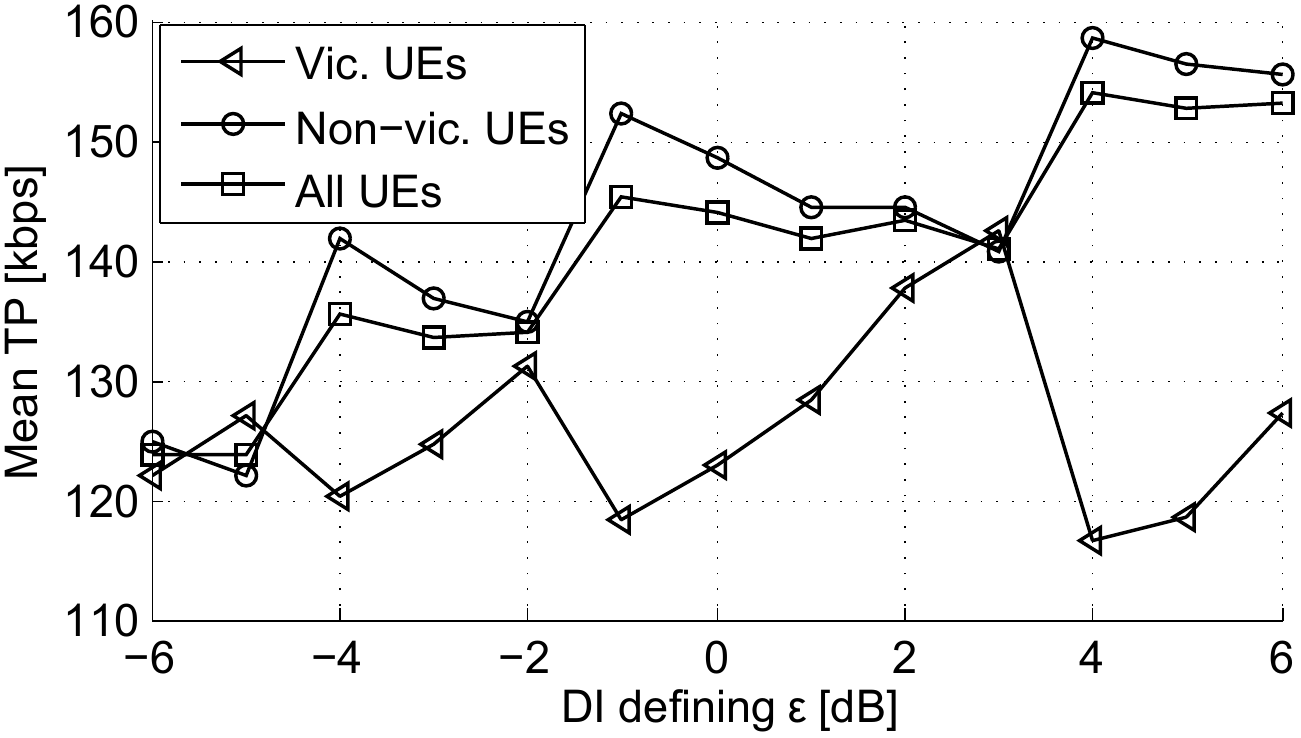}
    \caption{Mean round robin throughput of victim/non-victim/all MUEs in macro/femto scenario versus definition of dominant interferer.}\label{fig_mf_rr_k}
  \end{center}
\end{figure}

In Fig. \ref{fig_mf_rr_k} we present impact of the DI-defining coefficient $\kDI$ to MUE throughput via the same $\mf$ factor as in Subsection \ref{ssec:numAbsfMF}. Because differences in CDF curves are too small to notice with eyes, we put measured mean throughput on the y-axis. In an interval with the same number of ABSFs increasing value of $\mf$ leads to less victim and more non-victim MUEs per MBS, therefore victim MUE mean throughput is increasing, while non-victim MUE throughput is decreasing. For our parameters a good point of operation is with $\mf \!\in\! \left( 2\text{dB},3\text{dB} \right)$, where the number of ABSFs is lower than around $\mf \!=\! -2\text{dB}$ and the mean victim and non-victim MUE throughputs are relatively close to each other.

\begin{table}[t!]
  \caption{Victim MUE outage throughput and non-victim MUE mean throughput in kbps for different numbers of ABSFs in macro/femto scenario.}
  \label{tab_optDemo}
  \begin{center}
      \begin{tabular}{c||ccccc}
        \hline
        $\nAbsf$ & 1 & 2 & 3 & 4 & 5 \\
	    \hline
		Victim mean outage TP & 32 & 43.6 & 53.8 & 65.3 & 75.4 \\
		\hline
		Non-victim mean TP & 188.2 & 170.7 & 148.7 & 128.1 & 107.7 \\
		\hline
	  \end{tabular}
  \end{center}
\end{table}  

To justify the simplification of optimization problem \eqref{eq_opt1} into \eqref{eq_opt2} we present in Table \ref{tab_optDemo} mean outage throughput of victim MUEs and mean throughput of non-victim MUEs for different numbers of ABSFs $\nAbsf$. Mean throughput of non-victim MUEs is a clearly decreasing function of ABSFs, the simplification is thus well justified. Although the table shows that 2 ABSFs would suffice to fulfill $C_\text{V,min} \!=\! 40\text{kbps}$ requirement, this is because our simulation area is not large enough as mentioned at the beginning of this section.

\subsection{Macro/pico scenario}

\begin{figure}[t!]
  \begin{center}
    \includegraphics[scale=0.5]{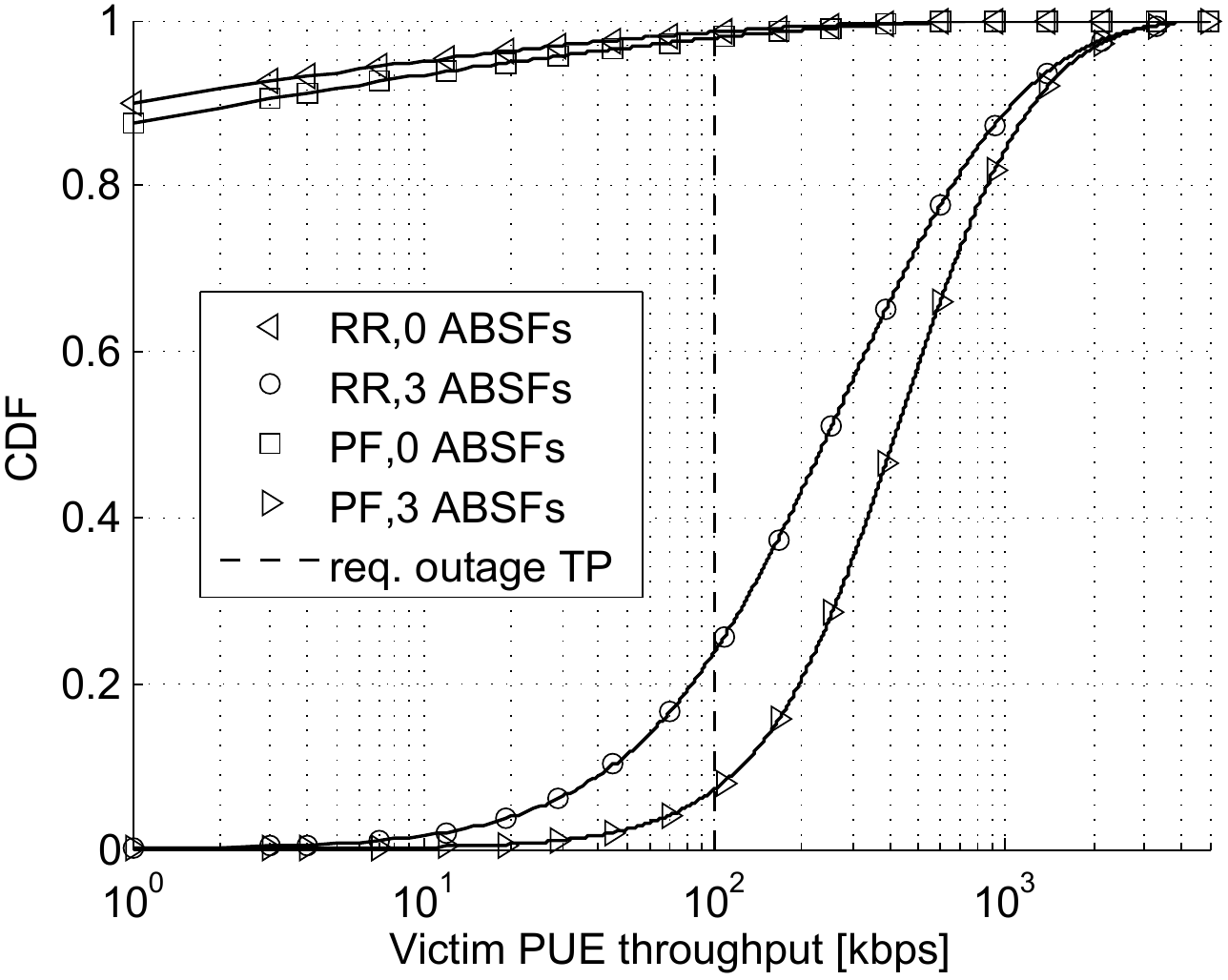}
    \caption{Throughput CDF of victim PUEs in macro/pico scenario with round robin and proportional fair scheduling, with and without ABSFs. The minimum average outage throughput is marked by a dashed line.}\label{fig_mp_tpCdf_vic}
  \end{center}
\end{figure}

In the second scenario we set the $C_\text{V,min}$ value to $100\text{kbits/s}$. In Fig. \ref{fig_mp_tpCdf_vic} we present throughput CDFs of victim PUEs with the same round robin and proportionally fair scheduling algorithms. As before, victim PUEs are scheduled in both ABSFs and NSFs, while non-victim PUEs have access only to NSFs. Without ABSFs, the victim PUEs experience very bad performance. Strong cross-layer interference (with $\pleM \!<\! \pleP$) with association bias $\cae$ on the top leads to very low throughput values. With both round robin and proportionally fair scheduler, more than 95\% of victim PUEs experience lower throughput than the $C_\text{V,min}$ requirement. With ABSFs, approximately 70\% and 90\% of victim PUEs reach the requirements using round robin and proportionally fair scheduler, respectively.

\begin{figure}[t!]
  \begin{center}
    \includegraphics[scale=0.5]{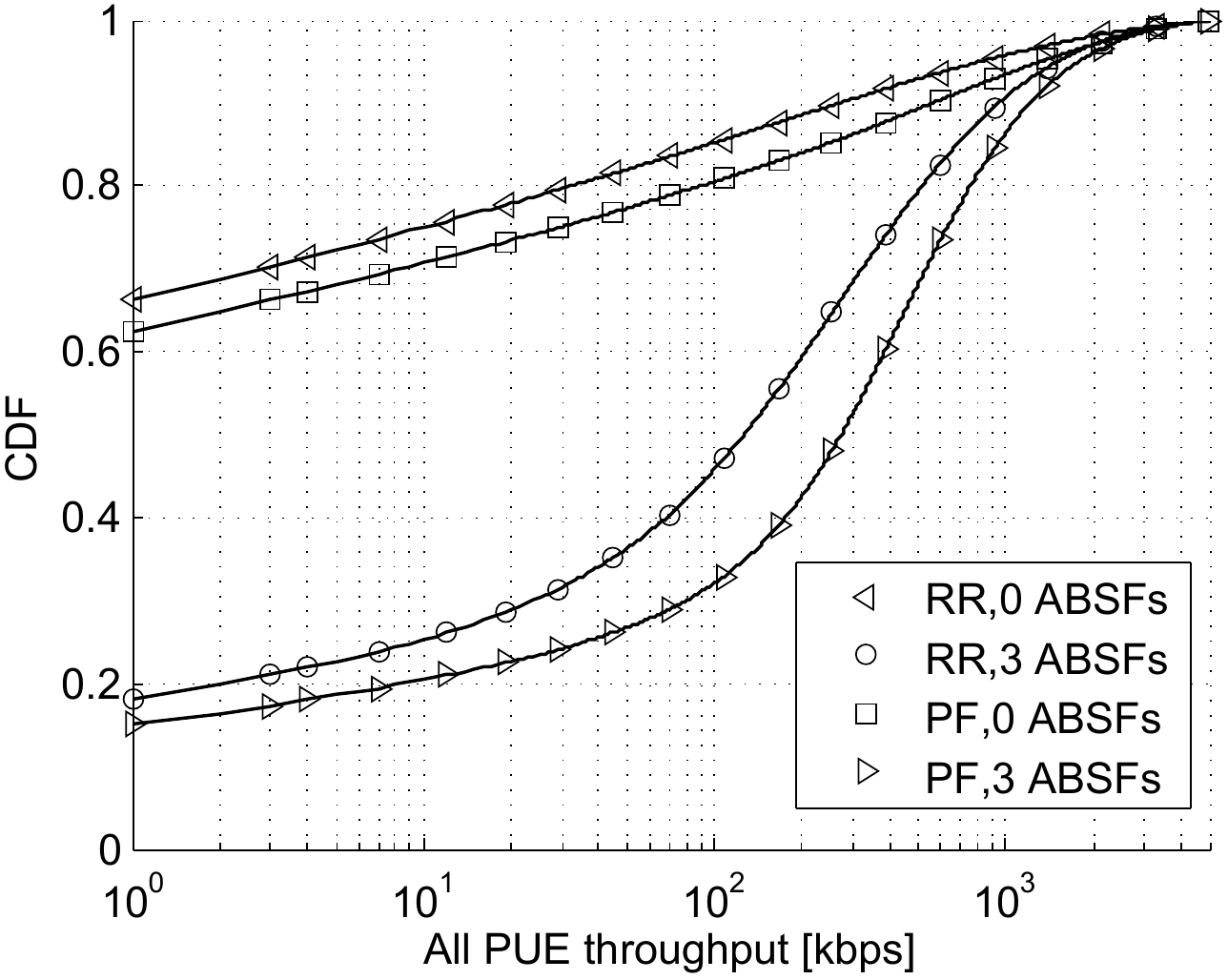}
    \caption{Throughput CDF of all PUEs in macro/pico scenario with round robin and proportional fair scheduling, with and without ABSFs.}\label{fig_mp_tpCdf_all}
  \end{center}
\end{figure}

\begin{figure}[t!]
  \begin{center}
    \includegraphics[scale=0.5]{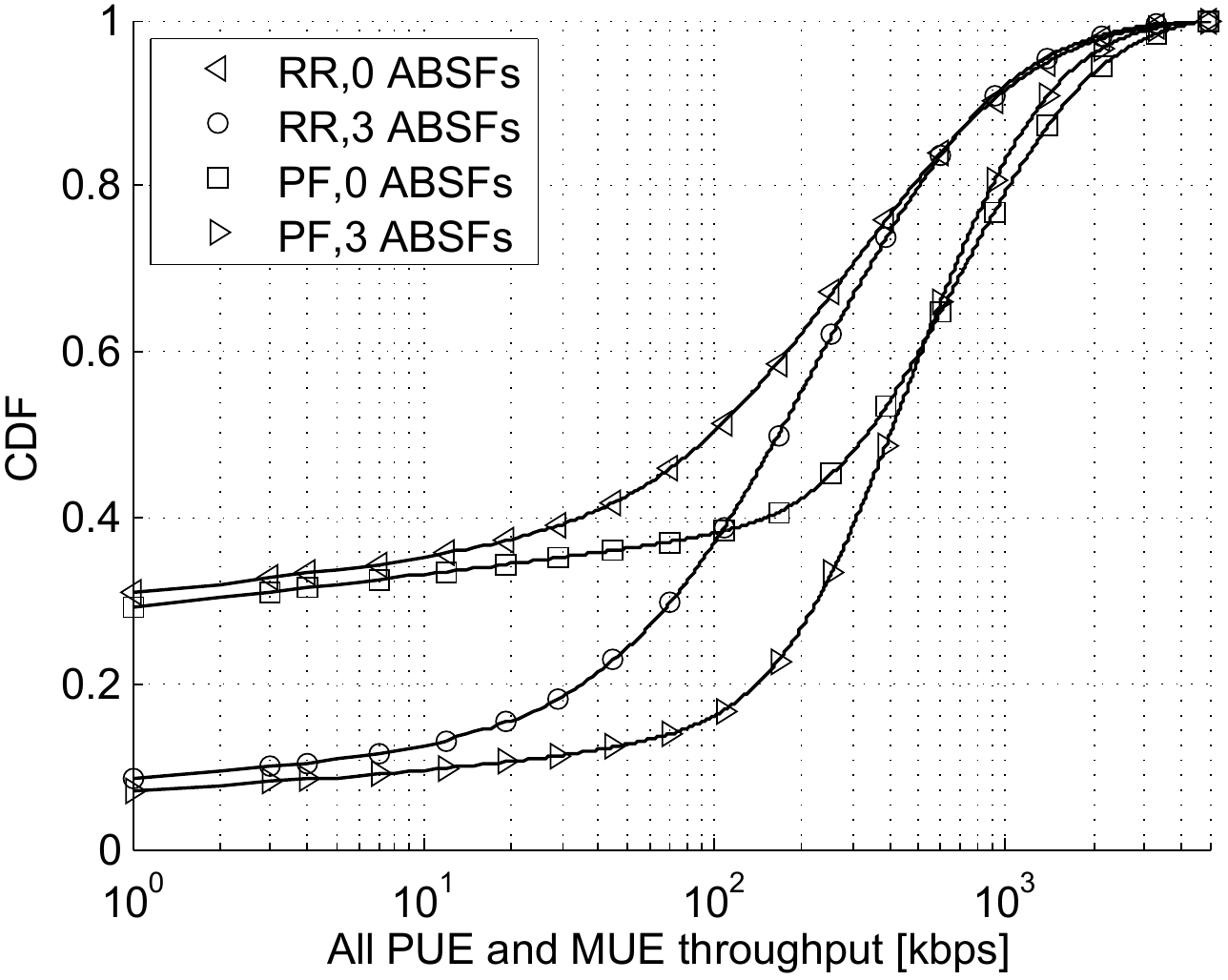}
    \caption{Throughput CDF of all MUEs and PUEs in macro/pico scenario with round robin and proportional fair scheduling, with and without ABSFs.}\label{fig_mp_tpCdf_ALL}
  \end{center}
\end{figure}

In Fig. \ref{fig_mp_tpCdf_all} we plot results from all PUEs. Because of the association bias the proportion of victim PUEs is quite high, therefore the decrease of non-victim PUE performance is not even visible in the figure. Overall, even though there are still PUEs experiencing zero performance, ABSFs substantially improve service in the pico layer. For completeness, Fig. \ref{fig_mp_tpCdf_ALL} also shows throughput CDFs for all UEs in the scenario, i.e., PUEs as well as MUEs. It confirms that ABSFs are very advantageous for given scenario - high performance gains in the lower percentiles are balanced by relatively small performance decrease in the higher end of the curve.

\begin{figure}[t!]
  \begin{center}
    \includegraphics[scale=0.5]{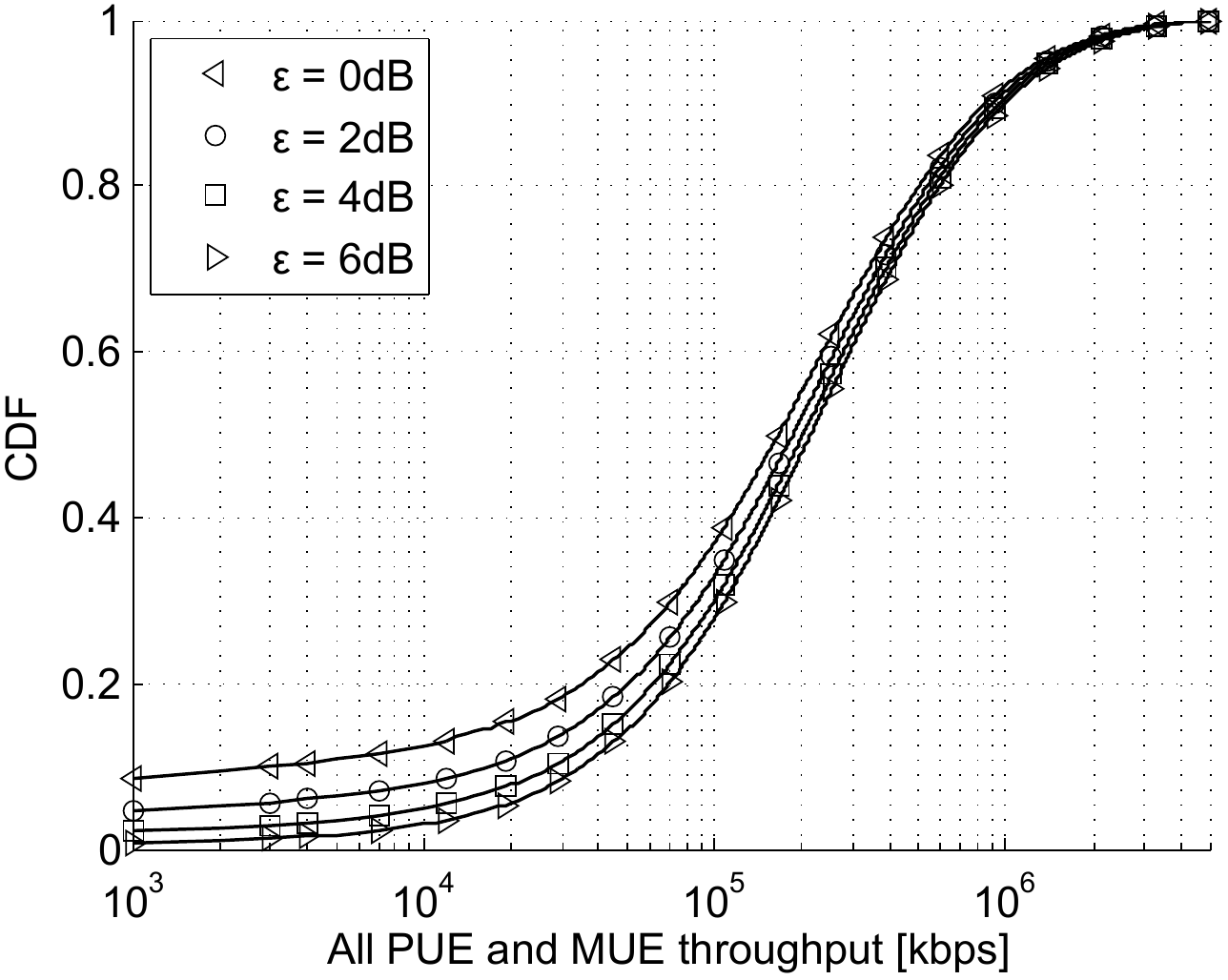}
    \caption{Throughput CDF of all MUEs and PUEs in macro/pico scenario with ABSFs, round robin scheduling and several values of DI-defining $\kDI_2$ via~$\mf$.}\label{fig_mp_tpCdf_ALL_k}
  \end{center}
\end{figure}

A sharp reader looking at Fig. \ref{fig_mp_tpCdf_all} and Fig. \ref{fig_mp_tpCdf_ALL} notices that despite using ABSFs, a percentage of PUEs is experiencing system outage, i.e., no throughput at all. These are PUEs that did not fulfill the DI definition via reference $\kDI_2$. To further investigate this important design property we plot in Fig. \ref{fig_mp_tpCdf_ALL_k} throughput CDFs of all UEs in macro/pico scenario with multiple values of $\mf$ that defines $\kDI_2 \!=\! \left[ \pP / \left( \mf \pM \right) \right]^{2 / \left( \pleP \!+\! \pleM \right)}$. Decreasing the DI definition threshold noticeably increases throughput values in the lower percentiles and thus reduces system outage. Because our reference scenario settings yield relatively low amount of PUEs per PBS, increasing the number of victim PUEs actually does not show any penalty in this result set. This will however not hold in general, as the PBSs are expected to be operational especially in areas with higher UE density.

\section{Conclusions}
\label{sec:outro}

Almost blank subframes (ABSFs) offer a simple and efficient way of decreasing the level of background cross-tier interference and thus give an opportunity to serve vulnerable users. We propose a way to approximate the required number of ABSFs based on Poisson point process network deployment statistics. We derive the necessary number of ABSFs as a formula that is easy to evaluate for macro/femto scenario with closed subscriber groups and macro/pico scenario with cell range expansion. We analyze dependence of the result on individual parameters, showing that while in macro/femto scenario the white residue interference in ABSF can be tolerated well, in macro/pico scenario its effect on the required number of ABSFs is substantial. Throughput simulations show that in macro/femto scenario the performance gain when using ABSFs is rather moderate and the system may work better with a more individual treatment of high FBS interference. On the other hand, in the macro/pico scenario we can see that, especially due to the association bias, using ABSFs improves the performance of the system considerably.

The fraction of victim users in outage decreases from over 95\% to 30\% or even 10\%, depending on the scheduling algorithm. Looking at all users in the macro/pico scenario, the small decrease in high throughput percentiles is more than balanced by substantial performance increase in the lower region.

\appendix[Derivation of round robin resource fractions]
\label{sec:appOmega}

\begin{figure*}[!t]
  \footnotesize
  \setcounter{MYtempeqncnt}{\value{equation}}
  \setcounter{equation}{59}
  \begin{equation}
    \label{eqBiasedVoronoiPdf}
    f_\text{m/f}^{(N)}(x) \approx \frac{d}{dx} F_S(x,p_\text{m/f}^{(\text{V})}) = \frac{ 343 e^{-x \left( \frac{7 \intM}{2} + S_1 \right)} \left( 1-e^{-x S_1} \right)
      \sqrt{\frac{7}{2\pi}} x^{5/2} \intM^{7/2} S_2^3 }
      { 5145 \sqrt{7} \kDI^6 \intF^3 \intM^3 S_3 + 15435 \sqrt{7} \kDI^4 \intF^2 \intM^4 S_3 + 15435 \sqrt{7} \kDI^2 \intF \intM^5 S_3 + 5145 \sqrt{7} \intM^6 S_3 - S_2^3 }
  \end{equation}
  \begin{equation}
    \label{eq:macroFemtoNSFomega}
    \overline{\Omega}_\text{m/f}^{(\text{N})} \approx \frac{ \intM^{9/2} \left( \frac{\sqrt{7}}{\intM^{7/2}} - \frac{2401}{S_0^{7/2}}
      - 2401 \left( \frac{\kDI^2 \intF + \intM}{S_2} \right)^{7/2} + 2401 \left( \frac{\kDI^2 \intF + \intM}{\intM S_0 + \kDI^2 \intF \left( 7 \intM + 4 \intMUE \right)} \right)^{7/2} \right) }
      { \sqrt{7} \intMUE - 16807 \intMUE \left( \frac{\intM}{7 \intM + 2 S_1} \right)^{9/2} }
  \end{equation}
  \begin{equation}
    \label{eq:macroFemtoABSFomega}
    \overline{\Omega}_\text{m/f}^{(\text{A})} \approx \frac{1}{\intM} \left( \frac{15 \sqrt{2\pi / 7}}{343 S_4^{7/2}} - \frac{30 \sqrt{2\pi}}{\left( 7 S_4 + 2 \intMUE \right)^{7/2}} 
      + \frac{15 \sqrt{2\pi}}{\left( 7 S_4 + 4 \intMUE \right)^{7/2}}\right)
      \left( \frac{15 \sqrt{2\pi / 7}}{343 S_4^{9/2}} - \frac{105 \sqrt{2\pi}}{\left( 7 S_4 + 2 \intMUE \right)^{9/2}} \right)^{-1}
  \end{equation}
  \setcounter{equation}{\value{MYtempeqncnt}}
  \hrulefill
  \vspace*{4pt}
\end{figure*}
\begin{figure}[t!]
  \begin{center}
    \includegraphics[scale=0.5]{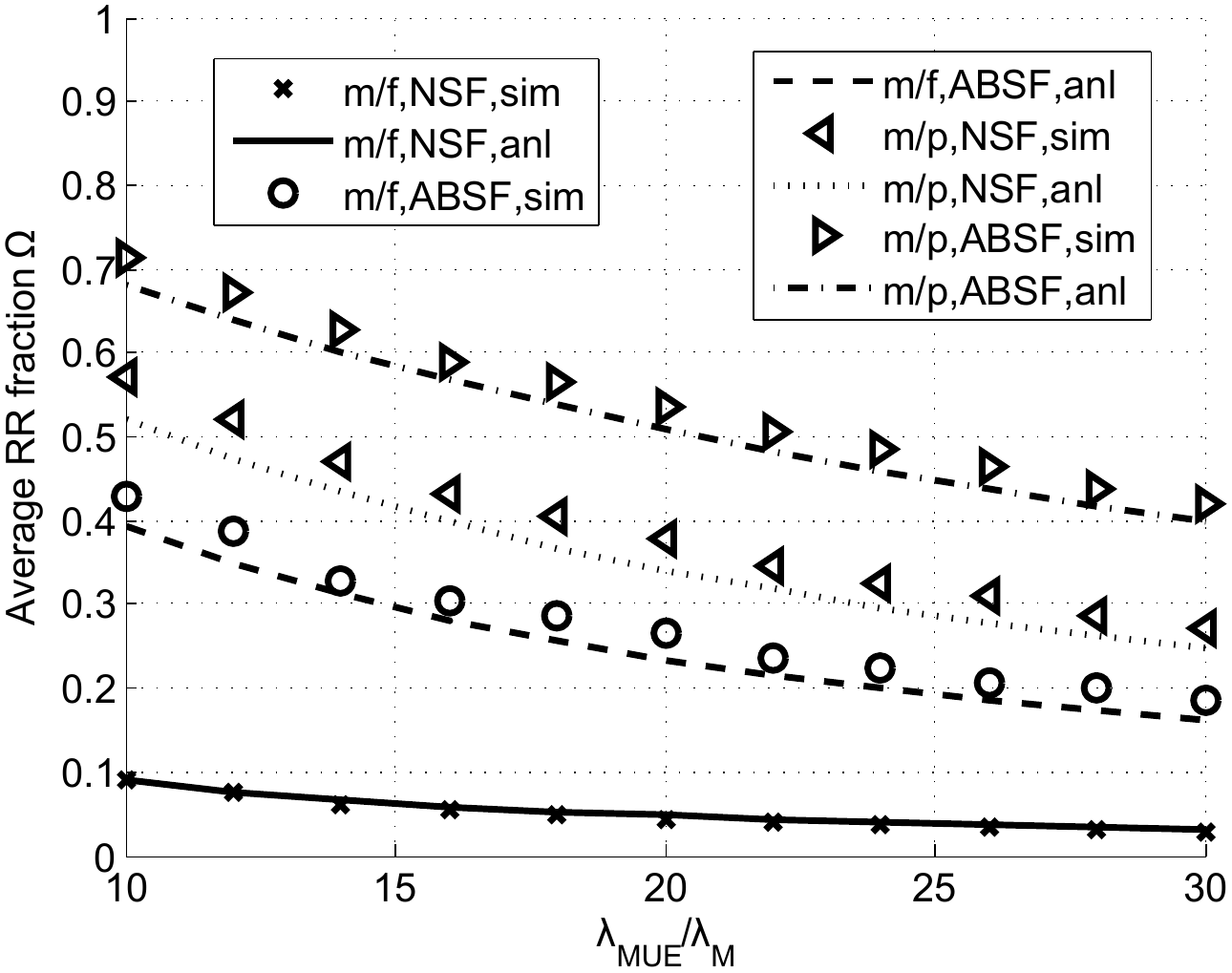}
    \caption{Average round robin fraction $\overline{\Omega}$ in macro/femto (m/f) and macro/pico (m/p) scenario. Markers represent simulated values, lines represent analytically approximated values.}\label{fig_RRfractionPrecision}
  \end{center}
\end{figure}

If a loaded BS governs $L$ associated UEs and all UEs have uniform traffic requirements, each UE will asymptotically be scheduled in $1/L$ fraction of available resource blocks. We are interested in expected value of this fraction for victim UEs. Using a clever approach from \cite{YuKi2012} we can write
\addtocounter{equation}{3}
\begin{align}
  \overline{\Omega} &= C_\Omega^\prime \sum_{\ell \!=\! 1}^\infty \frac{1}{\ell} \pr \left\{ \Omega \!=\! \frac{1}{\ell} \right\} \\
  &= C_\Omega \sum_{\ell \!=\! 1}^\infty \frac{1}{\ell} \ell \pr \left\{ L \!=\! \ell \right\} \\
  &= C_\Omega \sum_{\ell \!=\! 1}^\infty \int_0^\infty \pr \left\{ \left. L \!=\! \ell \right| A \!=\! x \right\} f_A(x) dx \\
  &= C_\Omega \sum_{\ell \!=\! 1}^\infty \int_0^\infty \frac{\left( x \intMUE \right)^\ell}{\ell !} e^{-\intMUE x} f_A(x) dx \\
  &= C_\Omega \int_0^\infty e^{-\intMUE x} f_A(x) \sum_{\ell \!=\! 1}^\infty \frac{\left( x \intMUE \right)^\ell}{\ell !} dx \\
  &= C_\Omega \int_0^\infty \left( 1 - e^{-\intMUE x} \right) f_A(x) dx,
\end{align}
where $f_A(x)$ is a PDF of an area where the given set of UEs can be located and $C_\Omega^\prime$ and $C_\Omega$ are normalization constants. We denote the area PDF $f_A(x)$ as $f_\text{m/f}^{(N)}(x)$ for all MUEs in macro/femto scenario (round robin fraction in NSF), $f_\text{m/f}^{(A)}(x)$ for victim MUEs in macro/femto scenario (round robin fraction in ABSF), $f_\text{m/p}^{(N)}(x)$ for all PUEs in macro/pico scenario and $f_\text{m/p}^{(A)}(x)$ for victim PUEs in macro/pico scenario. The normalization constant $C_\Omega^\prime$ is needed because we are excluding cases with no UEs associated to BS, while $C_\Omega$ takes also into account multiplying the PMF $\pr \left\{ L \!=\! \ell \right\}$ with a weight factor $\ell$ and can be calculated from
\begin{align}
  C_\Omega^{-1} &= \sum_{\ell \!=\! 1}^\infty \ell \pr \left\{ L \!=\! \ell \right\} \\
  &\vdots \nonumber \\
  &= \int_0^\infty \intMUE x f_A(x) dx.
\end{align}
In macro/femto scenario the PDF $f_\text{m/f}^{(N)}(x)$ can be obtained from the general approximation of a Poisson Voronoi cell area that has been found in \cite{FeNe2007} (and used in \cite{YuKi2012}) by adding a condition that there is at least one victim MUE present. The general Voronoi area for our macro PPP is drawn from PDF 
\begin{equation}
  f_{S}(x) \approx \intM \frac{343}{15} \sqrt{\frac{7}{2 \pi}} \left( \intM x \right)^\frac{5}{2} \exp \left( -\frac{7}{2} \intM x \right).
\end{equation}
The PDF $f_\text{m/f}^{(N)}(x)$ can be obtained from CDF
\begin{align}
  F_S(x,p^{(\text{V})}) &= \pr \left\{ \left. S \!\leq\! x \right| L_\text{V} \!>\! 0 \right\} \\
  &= \frac{ \pr \left\{ S \!\leq\! x , L_\text{V} \!>\! 0 \right\} }{ \pr \left\{ L_\text{V} \!>\! 0 \right\} } \\
  &= \frac{ \int_0^x \pr \left\{ \left. L_\text{V} \!>\! 0 \right| S \!=\! u \right\} f_S(u) du }
    { \int_0^\infty \pr \left\{ \left. L_\text{V} \!>\! 0 \right| S \!=\! u \right\} f_S(x) dx } \\
  &= \frac{ \int_0^x \left( 1 - \pr \left\{ \left. L_\text{V} \!=\! 0 \right| S \!=\! u \right\} \right) f_S(u) du }
    { \int_0^\infty \left( 1 - \pr \left\{ \left. L_\text{V} \!=\! 0 \right| S \!=\! u \right\} \right) f_S(x) dx } \\
  &= \frac{ \int_0^x \left( 1 - e^{-p^{(\text{V})} \intMUE u} \right) f_S(u) du }
    { \int_0^\infty \left( 1 - e^{-p^{(\text{V})} \intMUE x} \right) f_S(x) dx },\label{eq:voronoiCdf}
\end{align}
where $p^{(\text{V})}$ is a probability that a UE is a victim UE, i.e., in macro/femto scenario $p_\text{m/f}^{(\text{V})} \!=\! \pr \left\{ N_\text{DI}^{(\text{F})} \geq 1 \right\}$ from \eqref{eq_pr_onePlusDI}. After calculating the integrals in \eqref{eq:voronoiCdf} and differentiating the CDF we get a result \eqref{eqBiasedVoronoiPdf} with subterms:
\begin{equation}
  S_0 = 7 \intM + 2 \intMUE
\end{equation}
\begin{equation}
  S_1 = \frac{\kDI^2 \intF \intMUE}{\kDI^2 \intF + \intM}
\end{equation}
\begin{equation}
  S_2 = \left( 7 \intM^2 + \kDI^2 \intF S_0 \right)
\end{equation}
\begin{equation}
  S_3 = \sqrt{ \frac{ \intM \left( \kDI^2 \intF + \intM \right) }{ 7 \intM^2 + \kDI^2 \intF S_0 } }
\end{equation}
The PDF $f_\text{m/f}^{(A)}(x)$ of area where victim MUEs can be located is not trivial to find. The most sensible approach we came up with is transforming the unconditional Voronoi cell random variable using again the probability that MUE is a victim MUE, i.e.,
\begin{equation}
  \label{eq:macroFemtoVictimVoronoi}
  f_\text{m/f}^{(A)}(x) = \frac{1}{p_\text{m/f}^{(\text{V})}} f_\text{m/f}^{(N)} \left( \frac{x}{p_\text{m/f}^{(\text{V})}} \right).
\end{equation}
In macro/pico scenario the pdf of area of all UEs within a macrocell conditioned on presence of a victim PUE is
\begin{equation}
  f_\text{m/p}(x) = \frac{d}{dx} F_S(x,p_\text{m/p}^{(\text{V})}),
\end{equation}
where $p_\text{m/p}^{(\text{V})} \!=\! \pr \left\{ k_2 \rM < \rP < k_1 \rM \right\}$ as in \eqref{eq_pr_vicPUE}. For PUEs and victim PUEs we make a similar approximation as in \eqref{eq:macroFemtoVictimVoronoi} with addition of further multiplying the transformation coefficients with the average number of PBSs within MBS coverage, resulting in
\begin{equation}
  \begin{split}
  f_\text{m/p}^{(N)}(x) & = \frac{\intM}{\intP} \frac{1}{\pr \left\{ \rP \!<\! k_1 \rM \right\}} \\
  & \quad \times f_\text{m/p} \left(  \frac{\intM}{\intP} \frac{x}{\pr \left\{ \rP \!<\! k_1 \rM \right\}} \right),
  \end{split}
\end{equation}
where
\begin{equation}
  \pr \left\{ \rP \!<\! k_1 \rM \right\} = \frac{\kDI_1^2 \intP}{\kDI_1^2 \intP + \intM}
\end{equation}
and
\begin{equation}
  f_\text{m/p}^{(A)}(x) = \frac{\intM}{\intP} \frac{1}{p_\text{m/p}^{(\text{V})}} f_\text{m/p} \left( \frac{\intM}{\intP} \frac{x}{p_\text{m/p}^{(\text{V})}} \right).
\end{equation}
The round robin fraction values $\overline{\Omega}$ can be obtained in closed form for all cases. For macro/femto scenario we present them in \eqref{eq:macroFemtoNSFomega} and \eqref{eq:macroFemtoABSFomega} with subterms given here:
\begin{equation}
  S_4 = \intM \left( 1 + \frac{\intM}{\kDI^2 \intF} \right)
\end{equation}
Formulas for $f_\text{m/p}^{(\text{N})}(x)$, $f_\text{m/p}^{(\text{A})}(x)$, $\overline{\Omega}_\text{m/p}^{(\text{N})}$ and $\overline{\Omega}_\text{m/p}^{(\text{A})}$ we leave out of the paper because they are very spacious and do not have enough added value by themselves. To illustrate the precision of our approximations we show the $\overline{\Omega}$ values as functions of $\intMUE / \intM$ in Fig. \ref{fig_RRfractionPrecision}. The approximation of $\overline{\Omega}_\text{m/f}^{(\text{N})}$ is very good. In other cases our derived formulas give consistently lower values than the simulations. The reason behind this is that when transforming the Voronoi area PDFs $f_\text{m/f}(x)$ and $f_\text{m/p}(x)$ we neglect that dangerous zones around FBSs and coverage areas of PBSs can overlap. Nevertheless, the derived values $\overline{\Omega}_\text{m/f}^{(\text{A})}$, $\overline{\Omega}_\text{m/p}^{(\text{N})}$ and $\overline{\Omega}_\text{m/p}^{(\text{A})}$ serve as a good lower bound and ensure that the required number of ABSFs as given in Section \ref{sec:numAbsfs} will not be too low. 

\end{document}